\documentclass[aps,prl,twocolumn,showpacs,10pt,superscriptaddress,preprintnumbers,nofootinbib]{revtex4-1}
\pdfoutput=1
\usepackage{amssymb,amsmath,mathtools}
\usepackage{graphicx}
\usepackage{subfigure}
\usepackage[margin=7pt,justification=centerlast]{caption}
 \usepackage{xspace}
\usepackage{hyperref}

\allowdisplaybreaks

\newcommand{\nnlojet}{NNLO\protect\scalebox{0.8}{JET}\xspace}
\DeclareRobustCommand{\LO}{\text{LO}\xspace}
\DeclareRobustCommand{\NLO}{\text{NLO}\xspace}
\DeclareRobustCommand{\NNLO}{\text{NNLO}\xspace}
\DeclareRobustCommand{\N}[1]{\ensuremath{\text{N}^{#1}}} 

\def\beq{\begin{equation}}
\def\eeq{\end{equation}}
\def\bsp#1\esp{\begin{split}#1\end{split}}
\newcommand{\be}{\begin{equation}}
\newcommand{\ee}{\end{equation}}
\newcommand{\bea}{\begin{eqnarray}}
\newcommand{\eea}{\end{eqnarray}}

\def\ksl{\not{\hbox{\kern-2.3pt $k$}}}

\def\alphas{\alpha_s}

\def\spa#1.#2{\left\langle#1\,#2\right\rangle}
\def\spb#1.#2{\left[#1\,#2\right]}
\def\lor#1.#2{\left(#1\,#2\right)}
\def\sand#1.#2.#3{%
\left\langle\smash{#1}{\vphantom1}^{-}\right|{#2}%
\left|\smash{#3}{\vphantom1}^{-}\right\rangle}

\def\d{\text{d}}

\begin{document}

\title{Transverse Mass Distribution and Charge Asymmetry\\ in W Boson Production to Third Order in QCD}
\preprint{IPPP/22/32, ZU-TH 19/22, P3H-22-055, KA-TP-15-2022, CERN-TH-2022-083}
\author{Xuan~Chen}
\email{xuan.chen@uzh.ch}
\affiliation{Institute for Theoretical Physics, Karlsruhe Institute of Technology, 76131 Karlsruhe, Germany}
\affiliation{Institute for Astroparticle Physics, Karlsruhe Institute of Technology, 76344 Eggenstein-Leopoldshafen, Germany}
\affiliation{Physik-Institut, Universit\"at Z\"urich, Winterthurerstrasse 190, CH-8057 Z\"urich, Switzerland}
\author{Thomas~Gehrmann}
\email{thomas.gehrmann@uzh.ch}
\affiliation{Physik-Institut, Universit\"at Z\"urich, Winterthurerstrasse 190, CH-8057 Z\"urich, Switzerland}
\author{Nigel~Glover}
\email{e.w.n.glover@durham.ac.uk}
\affiliation{
Institute for Particle Physics Phenomenology, Physics Department, Durham University, Durham, DH1 3LE, UK}
\author{Alexander~Huss}
\email{alexander.huss@cern.ch}
\affiliation{Theoretical Physics Department, CERN, 1211 Geneva 23, Switzerland}
\author{Tong-Zhi~Yang}
\email{toyang@physik.uzh.ch}
\affiliation{Physik-Institut, Universit\"at Z\"urich, Winterthurerstrasse 190, CH-8057 Z\"urich, Switzerland}
\author{Hua~Xing~Zhu}
\email{zhuhx@zju.edu.cn}
\affiliation{Zhejiang Institute of Modern Physics, Department of
  Physics, Zhejiang University, Hangzhou, 310027, China\vspace{0.5ex}}

\begin{abstract}
Charged gauge boson production  at hadron colliders is a fundamental benchmark for the extraction of electroweak parameters and the understanding of the proton structure. To enable precision phenomenology for this process, we compute the third-order (\N3\LO) QCD corrections to the 
rapidity distribution and charge asymmetry in W boson production and to the transverse mass distribution of its decay products.  
Our results display substantial QCD corrections in kinematic regions relevant for Tevatron and  LHC measurements. 
We compare the numerical magnitude of the \N3\LO corrections with uncertainties from electroweak input parameters and illustrate their potential impact on the determination of the W boson mass.
\end{abstract}

\maketitle


\section{Introduction}
\label{sec:introduction}

Charged electroweak~(EW) gauge bosons ${\mathrm W}^\pm$ are produced copiously through the charged-current Drell-Yan process at hadron colliders~\cite{Drell:1970wh}. Measurements of the inclusive and differential properties of W production play a central role in tests of the Standard Model~(SM) of particle physics and in the search for novel physics effects beyond it, allowing precision determinations of electroweak parameters such as the W boson mass and weak mixing angle, and of parton distribution functions~(PDFs) in the proton. In the past, measurements have been performed at the Fermilab Tevatron and the Large Hadron Collider~(LHC), substantially improving our knowledge about the SM~\cite{CDF:2012gpf,D0:2012kms,ATLAS:2017rzl,LHCb:2021bjt,CDF:2018cnj,ATLAS:2019fgb,ATLAS:2019zci,CMS:2020cph}.

Very recently, using a sample of approximately 4 million W bosons collected at the Tevatron,  the CDF Collaboration reported a new measurement of the W boson mass using template fits to the transverse mass distribution in W boson decays and the transverse momentum distribution of the decay leptons~\cite{CDF:2022hxs}.  The new measurement displays significant tension with the SM expectation and has an unprecedentedly small uncertainty of $\pm 9.4\,\text{MeV}$. The new result
calls for a careful assessment of the theoretical predictions for the charged-current Drell-Yan process and their associated uncertainties.

Due to their importance for SM measurements, precision predictions for Drell-Yan production have been among the first applications of perturbative 
QCD at next-to-leading order (NLO,~\cite{Altarelli:1979ub}), next-to-next-to-leading order~(NNLO,~\cite{Hamberg:1990np}), and the inclusive 
corrections were accomplished at third order (\N3\LO) accuracy in QCD~\cite{Duhr:2020sdp} recently. Fully differential NNLO QCD 
corrections~\cite{Anastasiou:2003ds,Melnikov:2006kv,Catani:2009sm,Catani:2010en}, including the kinematics of the decay leptons, have been available for a while
and are routinely used in the experimental analysis of Drell-Yan data. 

Significant efforts have also gone in the derivation of EW~\cite{Dittmaier:2001ay,Baur:2004ig,CarloniCalame:2007cd} and 
mixed QCD-EW corrections~\cite{Dittmaier:2015rxo,Dittmaier:2020vra,Behring:2020cqi,Buonocore:2021rxx,Behring:2021adr}  for W production and into the combination 
of fixed-order predictions with resummation of large logarithmic corrections~\cite{Balazs:1997xd,Bozzi:2010xn,Becher:2011xn,Bizon:2019zgf}.

In this letter we present for the first time differential predictions for W production at \N3\LO  in QCD, including the 
gauge boson rapidity distributions and associated charge asymmetry as well as  the transverse mass distribution.  
 Compared to NNLO, we find large \N3\LO corrections to the rapidity distributions with non-overlapping scale uncertainty bands. For the 
 normalized transverse mass distribution, which plays an important role in the W mass measurement, we observe a high perturbative stability 
 of the predictions from NNLO to \N3\LO. We quantify the impact of varying EW input parameters around the peak region 
 of the transverse mass distribution, finding substantial effects that could have 
 an important impact on future precision EW measurements using charged-current Drell-Yan production. 

\section{Methodology}
\label{sec:theoryinput}
In this letter, we calculate the differential cross section $\d\sigma_{\text{DY}}$, of charged-current production using the $q_T$-subtraction formalism~\cite{Catani:2007vq}:
\begin{equation}
\d\sigma_{\text{DY}}^{\N3\LO}=\mathcal{F}_{\text{DY}}^{\N3\LO}\otimes\d\sigma_{\text{DY}}^{\LO}|_{q_T < q_T^{\text{cut}}} + \d\sigma_{\text{DY+jet}}^{\NNLO}|_{q_T>q_T^{\text{cut}}},
\label{eq:master}
\end{equation}
where a slicing parameter $q_T^{\text{cut}}$ is introduced to separate unresolved and resolved contributions. Our study is based on an established framework at \N3\LO~\cite{Cieri:2018oms,Billis:2019vxg,Chen:2021vtu} which integrates the unresolved and resolved contributions in a computationally demanding manner. The independence of the results on the unphysical slicing parameter $q_T^{\text{cut}}$ serves as a strong check. 

The unresolved contribution $\mathcal{F}_{\text{DY}}^{\N3\LO}\otimes\d\sigma_{\text{DY}}^{\LO}|_{q_T < q_T^{\text{cut}}}$ denotes the fixed-order prediction for producing a W boson with transverse momentum $q_T$ less than $q_T^{\text{cut}}$ within Soft-Collinear Effective Theory (SCET)~\cite{Bauer:2000ew,Bauer:2000yr,Bauer:2001yt,Bauer:2002nz,Beneke:2002ph}. It can be expanded into logarithmic terms $\alpha_s^m \ln^n (q_T^{\text{cut}}/M_{\mathrm W})$ and constant terms.
All logarithmic terms can be predicted through to \N3\LO~\cite{Chen:2018pzu,Billis:2019vxg} using the rapidity renormalization group formalism~\cite{Chiu:2012ir}. The key ingredients to achieve \N3\LO accuracy for color-singlet production are the constant terms. They arise from the boundary conditions for the renormalization group equation, namely the rapidity-divergent transverse-momentum-dependent soft function~\cite{Li:2016ctv} and beam functions~\cite{Luo:2019szz,Ebert:2020yqt,Luo:2020epw} at three loops, as well as the massless QCD form factor~\cite{Baikov:2009bg,Lee:2010cga,Gehrmann:2010ue}.

The resolved contribution above $q_T^{\text{cut}}$ is computed using the \nnlojet code for charged-current Drell-Yan-plus-jet production at NNLO~\cite{Gehrmann-DeRidder:2017mvr,Gehrmann-DeRidder:2019avi}. It is fully differential at NNLO accuracy by employing the antenna subtraction method~\cite{hep-ph/0505111,hep-ph/0612257,1301.4693}. Sufficient numerical precision is mandatory to enable the cancellation between resolved and unresolved contributions at $q_T^{\text{cut}}$. This is achieved through dedicated optimization~\cite{Chen:2018pzu,Bizon:2019zgf,Chen:2021vtu} of phase space generation and subtraction terms to enable robust coverage in the unresolved regions for small values of the slicing parameter
$q_T^{\text{cut}}$. The cancellation of  $q_T^{\text{cut}}$-dependent terms between resolved and unresolved contributions in (\ref{eq:master}) is accurate only 
up to power-suppressed terms at $\mathcal{O}(\alpha_s^3(q_T^{\text{cut}}/M_{\mathrm W})^2)$ which are unaccounted for in the unresolved piece. These terms are found to be 
sufficiently suppressed to no longer affect the final result for $q_T^{\text{cut}}\sim1.5$ (0.75)\,GeV at the LHC (Tevatron), which is validated for each LHC (Tevatron) observable by varying $q_T^{\text{cut}}$ by $\pm 0.5$ (0.25)\,GeV.

The decay of  the W boson into a charged lepton and a neutrino is described at leading order with a  Breit-Wigner parametrisation of the W propagator using a fixed width.
To assess the impact of higher order QCD corrections and EW input parameters, we use the PDG~\cite{ParticleDataGroup:2020ssz} values $M_{\mathrm W}=80.379$\,GeV and $\Gamma_{\mathrm W}=2.085$\,GeV as the default setup and compare predictions with variations of $M_{\mathrm W}$ and $\Gamma_{\mathrm W}$.

\section{Results}
\label{sec:results}
Applying the $q_T$-subtraction method described above, we study charged-current Drell-Yan production at fully differential \N3\LO accuracy in proton-proton collisions with center-of-mass energy at 13\,TeV. We use the central member of \verb|NNPDF3.1| and \verb|NNPDF4.0| NNLO PDFs~\cite{NNPDF:2017mvq,NNPDF:2021njg} with $\alphas(M_{\mathrm Z})=0.118$ throughout the calculation and the scale evolution is performed with LHAPDF~\cite{Buckley:2014ana}. The electroweak couplings are determined using the 
$G_\mu$ scheme with: $M_{\mathrm Z}=91.1876$\,GeV, $\Gamma_{\mathrm Z}=2.4952$\,GeV, $G_F=1.1663787\times 10^{-5}$\,GeV$^{-2}$~\cite{ParticleDataGroup:2020ssz}. The CKM parameters are taken at their PDG values~\cite{ParticleDataGroup:2020ssz} in all Tevatron 
predictions, while a diagonal CKM matrix is used for LHC predictions. For absolute cross sections, the CKM effects are negligible for LHC energies (0.2\%) but relevant at 2\% level at Tevatron energies (largely due to the different partonic composition in proton-antiproton collisions). 
For normalized distributions without fiducial cuts, the CKM effects are negligible throughout.  
The central factorisation and renormalisation scales are chosen to be the invariant mass of final state leptons, $\mu_F=\mu_R=m_{\ell\nu}$. To estimate theoretical uncertainties, we adopt the 7-point scale variation of $\mu_F$ and $\mu_R$ by a factor of two while enforcing $1/2\le\mu_F/\mu_R\le2$.

\begin{figure}[t]
 \centering
\includegraphics[width=\columnwidth]{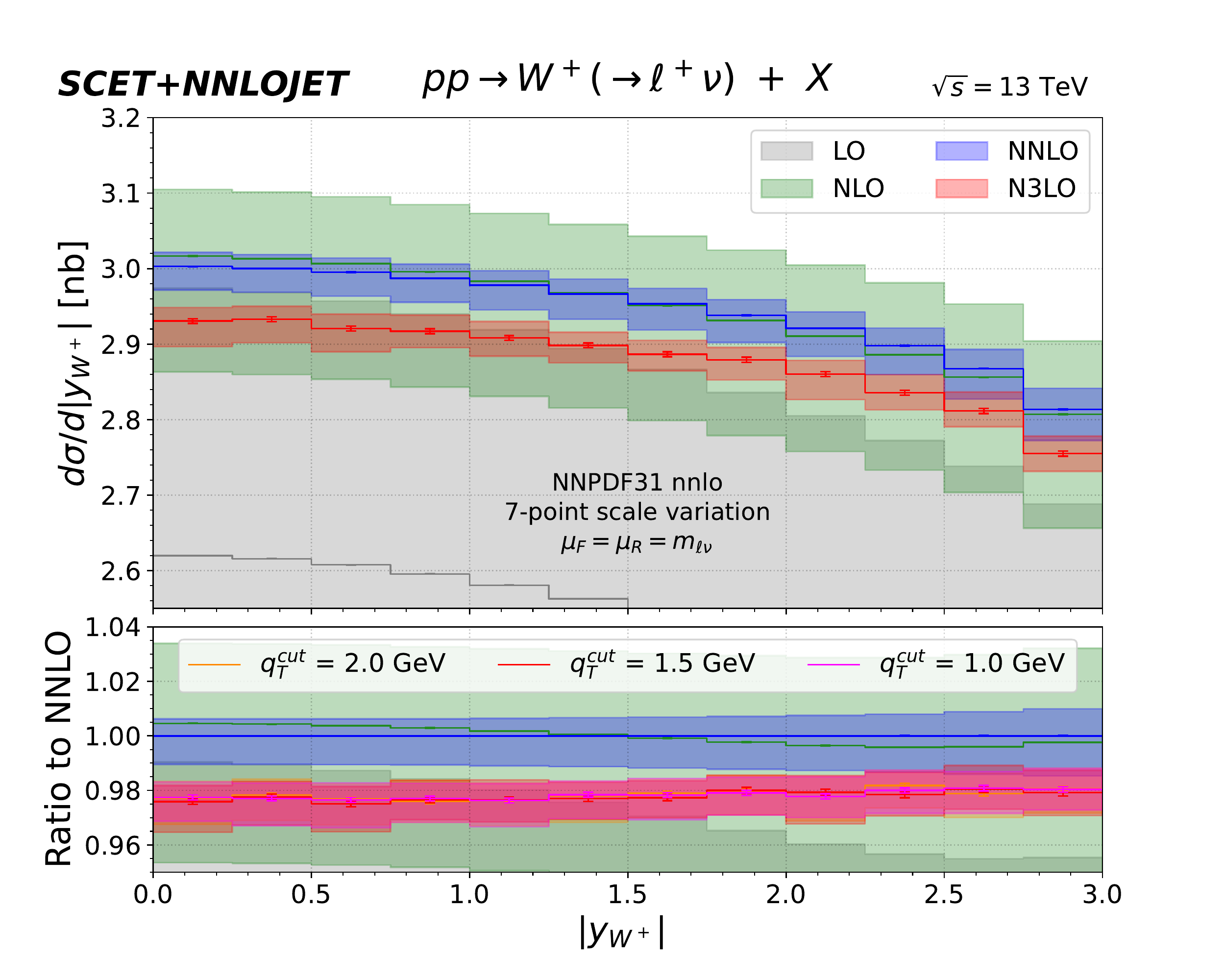}\\
\includegraphics[width=\columnwidth]{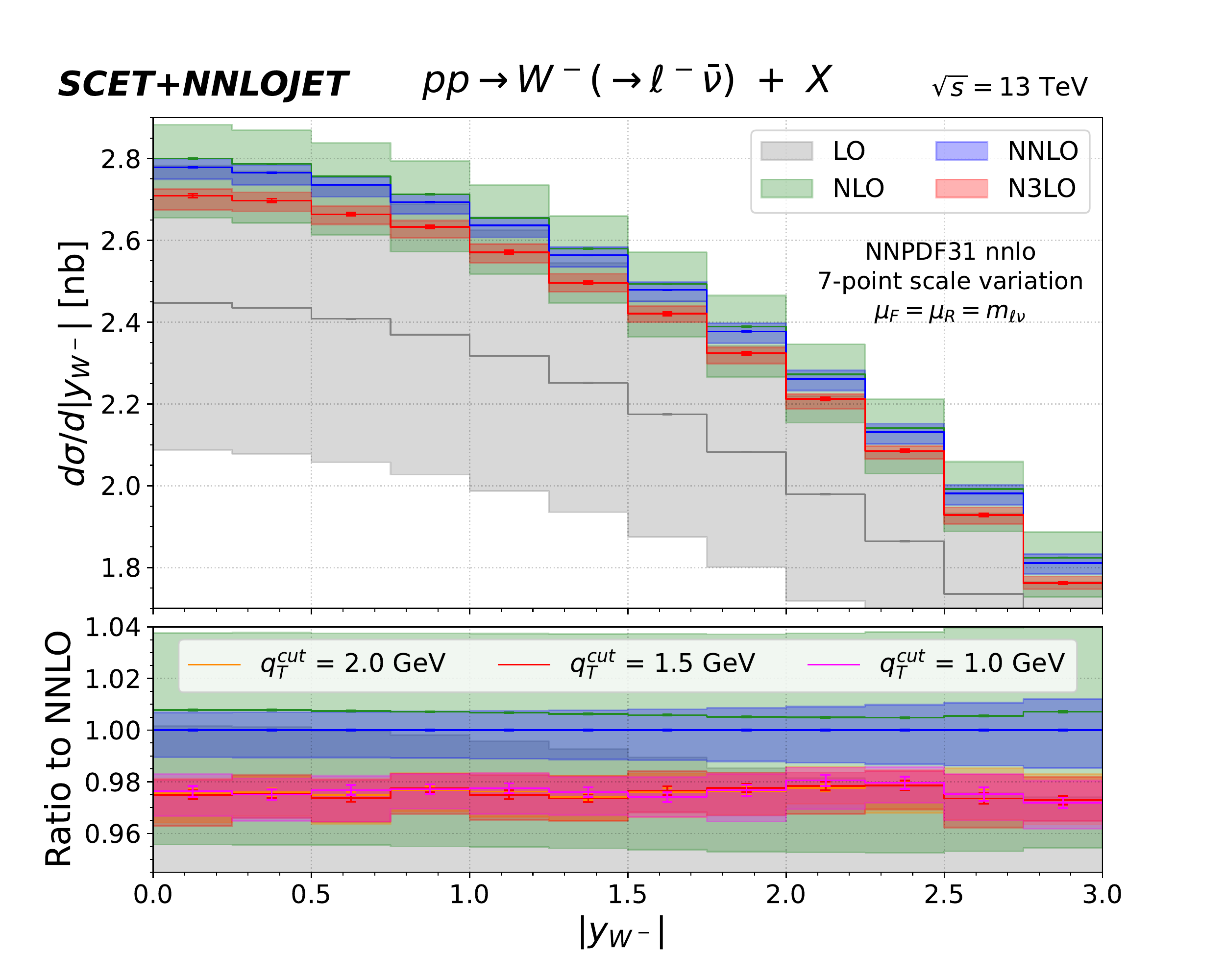}
\caption{W boson rapidity distributions from \LO to \N3\LO accuracy at the LHC. The colored bands represent theory uncertainties from 7-point scale variation. The bottom panels show 
the ratio with respect to NNLO, for three  different values of $q_T^{\rm cut}$.}
\label{fig:yW}
\end{figure}

In Fig.~\ref{fig:yW}, we show the rapidity distributions of the Drell-Yan pairs from ${\mathrm W}^+$ and ${\mathrm W}^-$ decays, with no fiducial cuts applied. Fixed order contributions with up to \N3\LO accuracy are included with the bottom panels showing their ratio with respect to the central NNLO result. The colored bands represent theory uncertainties from the 7-point scale variation and the error bars indicate the numerical integration error. Our state-of-the-art predictions at \N3\LO accuracy amount to a contribution of about $-2.5\%$ with respect to NNLO with relatively flat corrections for all rapidities. While the NLO and NNLO scale variation bands overlap, the \N3\LO prediction is found to be non-overlapping with the previous order within the respective
scale uncertainties. This feature at \N3\LO has already been observed for the total cross sections for neutral current~\cite{Duhr:2020seh,Duhr:2021vwj} and charged-current~\cite{Duhr:2020sdp} Drell-Yan production and for 
the  neutral-current Drell-Yan rapidity distribution~\cite{Chen:2021vtu} and fiducial cross sections~\cite{Chen:2022cgv}. 
 The relative size of scale variation remains comparable at NNLO and \N3\LO at about $\pm1\%$ for central rapidity and slightly increasing at large rapidity. We use three different $q_T^{\rm cut}$ values (1, 1.5 and 2\,GeV) to confirm the $q_T^{\rm cut}$-independence of the results within integration errors.
 A strong check on our results
 is provided by the rapidity-integrated charged current Drell-Yan cross section at  \N3\LO, where our results for   $q_T^{\rm cut}=1.5$~GeV agree 
 with~\cite{Duhr:2020sdp} within our numerical integration error of 1.5 per-mille. 
\begin{figure}[t]
 \centering
\includegraphics[width=\columnwidth]{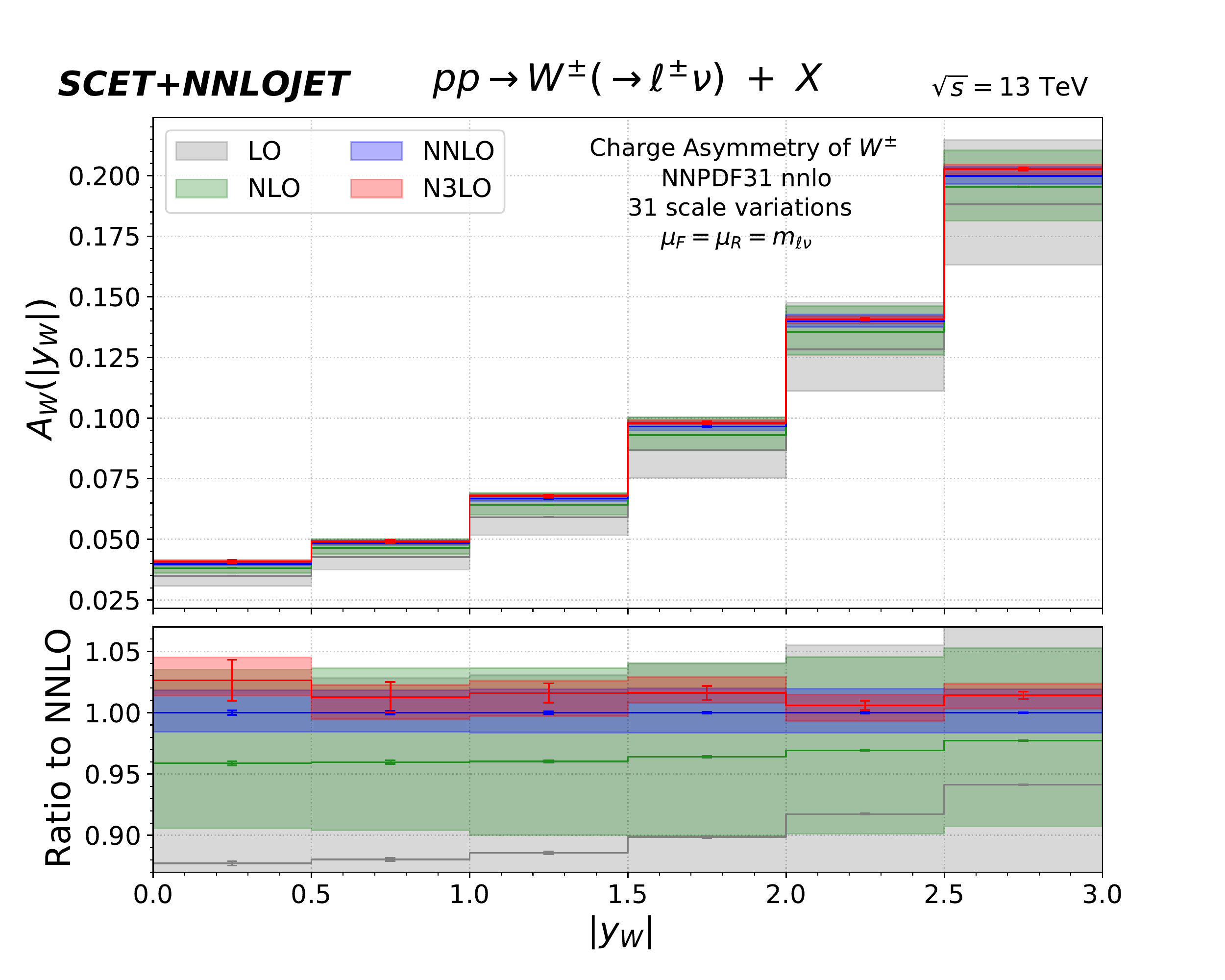}
\caption{W boson charge asymmetry distribution from \LO to \N3\LO at the LHC. The colored bands represent theory uncertainties from 31 scale variations. The bottom panel is
the ratio with respect to NNLO.}
\label{fig:CAS}
\end{figure}

The W boson charge asymmetry $A_{\mathrm W}$ at hadron colliders reveals details of the proton structure. It has been measured at the Tevatron~\cite{CDF:2009cjw,D0:2013lql} and the LHC~\cite{ATLAS:2011pph,ATLAS:2019fgb,CMS:2020cph}
and is defined as 
\begin{equation}
A_{\mathrm W}(|y_{\mathrm W}|)=\frac{\d\sigma/\d|y_{{\mathrm W}^+}|-\d\sigma/\d|y_{{\mathrm W}^-}|}{\d\sigma/\d|y_{{\mathrm W}^+}|+\d\sigma/\d|y_{{\mathrm W}^-}|}.
\label{eq:CAS}
\end{equation}
In Fig.~\ref{fig:CAS}, we display the predictions of $A_{\mathrm W}(|y_{\mathrm W}|)$ at 13\,TeV center of mass energy with up to \N3\LO corrections. We independently vary the scale choices between the numerator and the denominator of Eq.~\eqref{eq:CAS} while requiring $1/2\le\mu/\mu'\le2$ for any pair of scales, leading to 31 combinations. Their envelope is used to estimate the theoretical uncertainty. We observe positive \N3\LO corrections of about $2\%$ relative to the NNLO predictions. The \N3\LO contribution is
  not flat in rapidity. In contrast  to the individual rapidity distributions, the charge asymmetry  converges well from NLO to \N3\LO with scale variation uncertainty reduced to about $\pm 1.5\%$ at \N3\LO.

\begin{figure}[t]
 \centering
\includegraphics[width=\columnwidth]{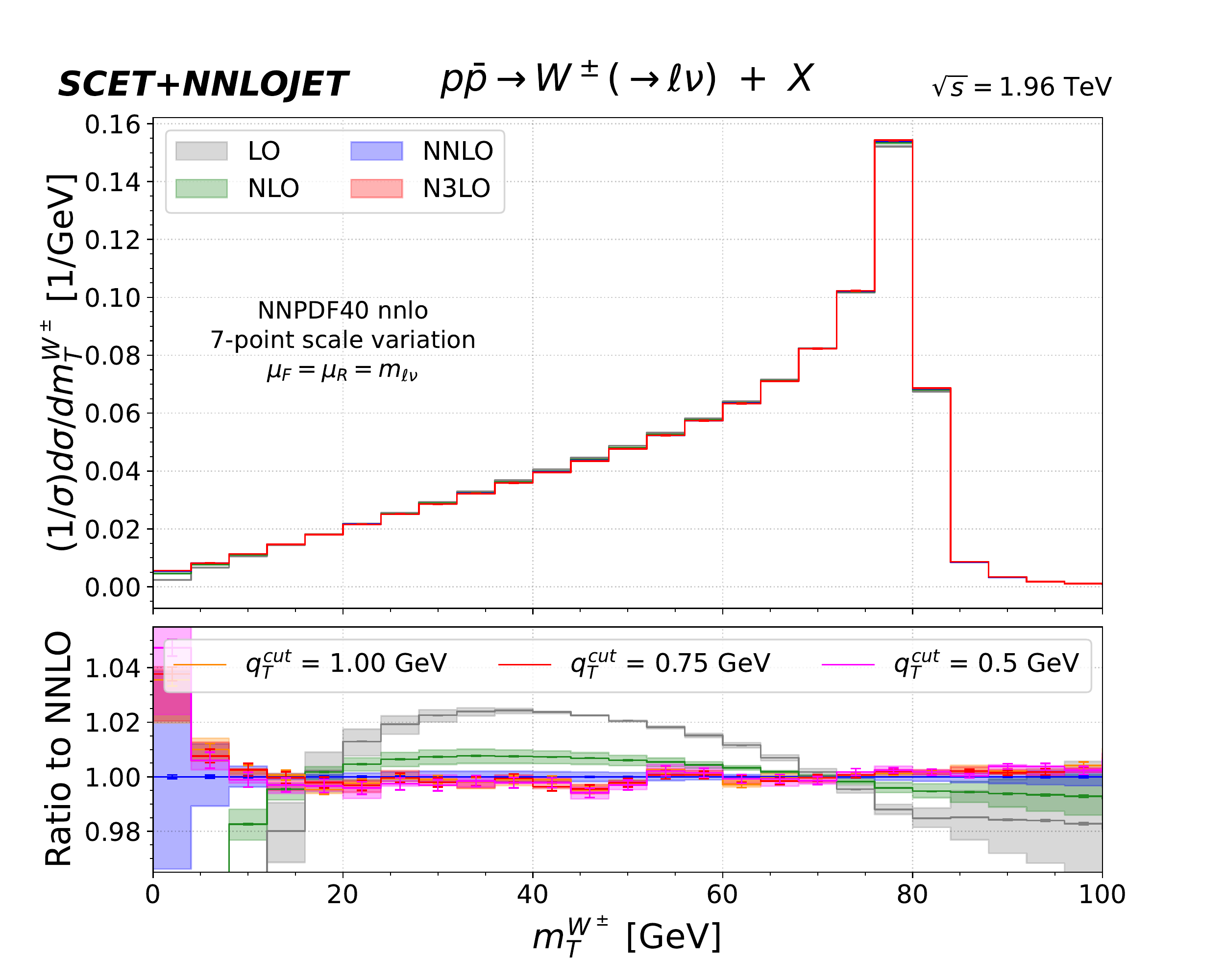}\\
\includegraphics[width=\columnwidth]{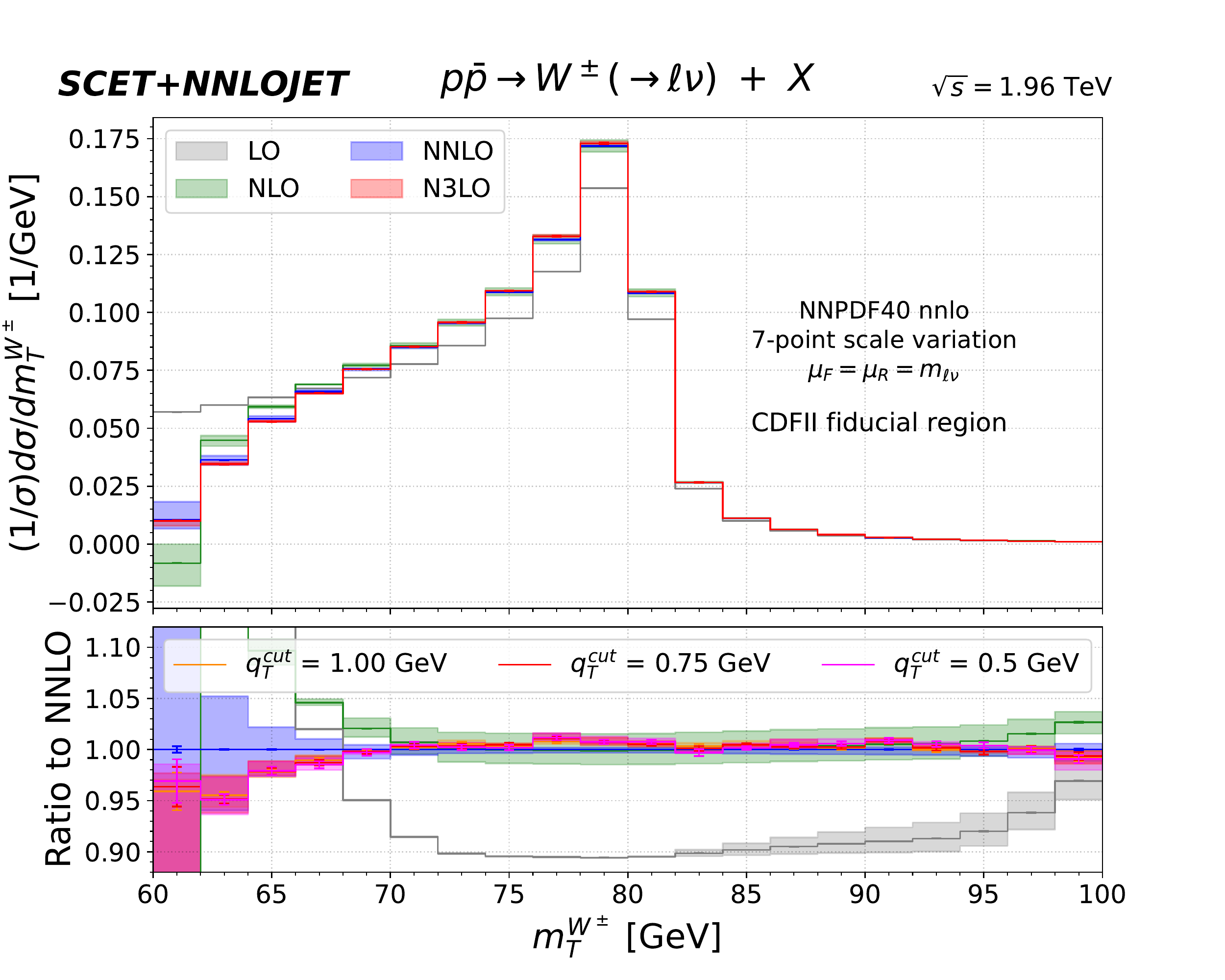}
\caption{Normalised W$^\pm$ transverse mass distribution from \LO to \N3\LO accuracy at the Tevatron without (upper) and with (lower) CDFII fiducial cuts. The colored bands represent theory uncertainties from 7-point scale variation. The bottom panel is
the ratio with respect to NNLO, with different cutoff $q_T^{\rm cut}$.}
\label{fig:mtFO}
\end{figure}

Finally, we consider the transverse mass distribution in charged-current Drell-Yan production. The transverse mass is constructed as 
\begin{equation}
m_T^{{\mathrm W}^{\pm}} = \sqrt{2E_T^{\ell^{\pm}}E_T^{\nu}(1-\text{cos}\Delta\phi)},
\end{equation}
with $E_T^{\ell^{\pm}(\nu)}$ denoting the transverse energies
of the final state charged lepton and neutrino and  $\Delta\phi$ being their azimuthal angle difference. 
It is a characteristic observable in measurements of $M_{\mathrm W}$~\cite{D0:2012kms,ATLAS:2017rzl,LHCb:2021bjt,CDF:2022hxs} and $\Gamma_{\mathrm W}$~\cite{CDF:2007tdb,D0:2009oet} at hadron colliders, since its distribution peaks around $M_{\mathrm W}$ and the shape of its tail is sensitive to $\Gamma_{\mathrm W}$. 
Precise predictions for the $m_T^{{\mathrm W}^{\pm}}$ distribution are vital for the measurement of W boson mass and width, which are 
based on fitting theory templates for the 
normalized distribution to data in the experimental analysis. The most recent measurement of $M_{\mathrm W}$ by CDFII collaboration reports $\pm 9.4$\,MeV overall uncertainty among which $\pm5.2$\,MeV arises from theoretical modelling~\cite{CDF:2022hxs}. The different sources of modelling uncertainties have 
subsequently been revisited~\cite{Isaacson:2022rts,Gao:2022wxk}, largely supporting the CDFII approach~\cite{CDF:2022hxs} while however not accounting 
for the state-of-the-art fixed-order predictions in mixed QCD-EW~\cite{Behring:2021adr} and fixed-order QCD corrections. 

Fig.~\ref{fig:mtFO} presents the normalized ${\mathrm W}^\pm$ boson transverse mass distribution at the Tevatron. With the newly computed \N3\LO corrections, 
it establishes a new state-of-the-art in the precise description of this observable. The inclusive distribution without fiducial cuts is displayed in the upper frame, while 
the fiducial cuts on charged lepton and neutrino of the CDFII analysis~\cite{CDF:2022hxs} are applied in the lower frame: $p_{T,l},E_{T,\nu} \in [30,55]$~GeV, 
$|\eta_l|< 1$ and $p_{T,W^\pm} < 15$~GeV. 
For the \N3\LO coefficient, we compensate the linear $q_T^{\text{cut}}$-dependence due to the fiducial cuts through a recoil prescription~\cite{Camarda:2021jsw,Buonocore:2021tke} where the unresolved contribution in Eq.~\eqref{eq:master} is active if all fiducial requirements are satisfied after boosting Born kinematics to finite $q_T$ below $q_T^{\text{cut}}$.
At \N3\LO, corrections are 
very uniform in the peak region for both inclusive and fiducial distribution, while displaying some kinematical dependence at low $m_T^{{\mathrm W}^{\pm}}$ below 12\,GeV in 
the inclusive distribution and below 68\,GeV in the fiducial distribution.
 Starting from NNLO, we observe a stabilization of scale uncertainties to  the level of $\pm1\%$.

\begin{figure}[t]
\centering
\includegraphics[width=\columnwidth]{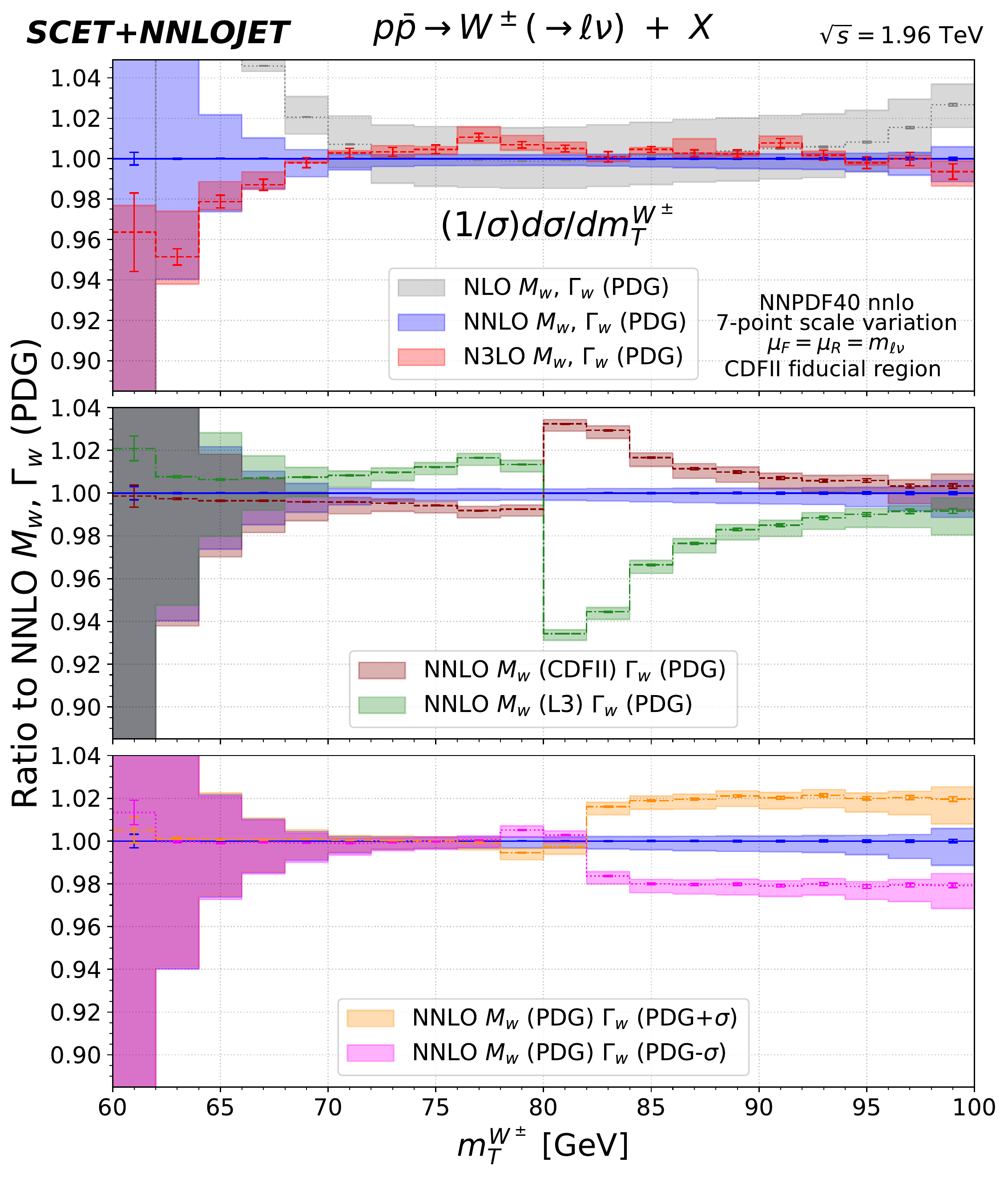}
\caption{Normalized ${\mathrm W}^{\pm}$ transverse mass distribution at the Tevatron with CDFII fiducial cuts. The \NLO to \N3\LO corrections are in the top panel, with different $M_{\mathrm W}$ values from PDG, CDFII and L3 in the middle panel and with different $\Gamma_{\mathrm W}$ with PDG central value and $\pm\ 1\sigma$ uncertainties in the bottom panel. All distributions are compared to the NNLO result with PDG central values. The colored bands represent theory uncertainties from 7-point scale variation.}
\label{fig:mtratio}
\end{figure}


Our newly derived predictions for the  $m_T^{{\mathrm W}^{\pm}}$ distribution allow us to compare different sources of theory uncertainty arising from higher perturbative orders and
from variations of the input parameters for $M_{\mathrm W}$ and $\Gamma_{\mathrm W}$ in Fig.~\ref{fig:mtratio}. For this purpose, we consider the PDG baseline values  $M_{\mathrm W}=80.379$\,GeV
and $\Gamma_{\mathrm W}=2.085$\,GeV, supplemented by the values of  $M_{\mathrm W}$ according to
the  measurement from CDFII (80.433\,GeV)~\cite{CDF:2022hxs} and L3 (80.27\,GeV)~\cite{L3:2005fft}, as well as a variation of $\Gamma_{\mathrm W}$ within the PDG
uncertainty of $\pm$42\,MeV~\cite{ParticleDataGroup:2020ssz}.

The top panel of Fig.~\ref{fig:mtratio} demonstrates that the normalized $m_T^{{\mathrm W}^\pm}$ fiducial distribution changes substantially in shape when going from
NLO to NNLO, but remains much more stable especially away from the peak region upon inclusion of the newly derived \N3\LO corrections. 
The sensitivity to the input parameters can thus be reliably quantified based on the NNLO predictions. The middle panel compares the
normalized $m_T^{{\mathrm W}^{\pm}}$ distributions for fixed  $\Gamma_{\mathrm W}=2.085$\,GeV and $M_{\mathrm W}$ values from  CDFII~\cite{CDF:2022hxs}
and L3~\cite{L3:2005fft}, thus quantifying the magnitude and shape of the resulting variations in the distribution.
As expected, a strong sensitivity on $M_{\mathrm W}$ around the peak region is observed. Compared to the change in the normalized distribution from NLO to NNLO, we observe that an NLO-based
template fit could experience a slight pull towards larger values  of $M_{\mathrm W}$ to compensate for missing NNLO corrections.  
 These variations are to be contrasted 
with the lower panel of Fig.~\ref{fig:mtratio}, where  $\Gamma_{\mathrm W}$ is varied by its PDG uncertainty of $\pm$42\,MeV for fixed PDG $M_{\mathrm W}$, which
basically affects the distributions only above the peak region, with a magnitude comparable to the effect of the $M_{\mathrm W}$ variation. 
 In contrast to the sensitivity on $\Gamma_{\mathrm W}$ of the CDFII measurement recently reported in~\cite{Isaacson:2022rts}, a more realistic 
 assessment of the uncertainties from  $\Gamma_{\mathrm W}$ may thus be warranted.

\section{Conclusions}
\label{sec:conclusion}
In this letter, we have produced state-of-the-art predictions for differential distributions in the charged-current Drell-Yan production to third order in perturbative QCD. We used the $q_T$-subtraction method at \N3\LO, by combining an NNLO calculation for the production of a charged Drell-Yan pair at large $q_T$ and leading-power factorised predictions from SCET at small $q_T$. The robust numerical quality of both contributions allowed us to consistently check the cancellation of the $q_T^{\text{cut}}$ dependence in our predictions. 

We presented differential distributions for the rapidity charge asymmetry at the LHC and for $m_T^{{\mathrm W}^\pm}$  at the Tevatron.
Our results display modest \N3\LO corrections in asymmetries and normalized distributions, that are usually within the uncertainties of the NNLO predictions. 
 \N3\LO perturbative uncertainties estimated by scale variations are found to be about $\pm1\%$ to $\pm 1.5\%$ throughout. 
 Distortions to the shape of the distributions are minimal at \N3\LO and only become visible outside the peak region of the $m_T^{{\mathrm W}^\pm}$ distribution. 

On the CDFII fiducial  $m_T^{{\mathrm W}^\pm}$ distribution~\cite{CDF:2022hxs}, we studied the impact of perturbative corrections and of variations of EW input parameters.
 We observed only minimal effects in going from NNLO to \N3\LO. Variations of $M_{\mathrm W}$ and $\Gamma_{\mathrm W}$ within their respective experimental uncertainties led on the other hand to 
characteristic shifts in the shape of the normalized transverse mass distributions at a level between $2\%$ to $6\%$ around the peak region. 

With the newly derived N3LO corrections, our results establish a new state-of-the-art for the perturbative description of W boson production at hadron colliders. They yield perturbative QCD uncertainties at the sub-per-cent level, which combine with recent results on QCD resummation, electroweak as well as mixed QCDxEW corrections to enable precision physics studies with upcoming LHC data.

\begin{acknowledgments}
\emph{Acknowledgements.} The authors would like to thank
Claude Duhr, Falko Dulat and Bernhard Mistlberger for
discussions and Julien Baglio for providing detailed results for the predictions in~\cite{Duhr:2020sdp}.
We are grateful to Aude Gehrmann-De Ridder, Tom Morgan and Duncan Walker for their contributions to the $V$+jet process in the \nnlojet code. This work has received funding from the Swiss National Science Foundation (SNF) under contract 200020-204200 and from the European Research Council (ERC) under the European Union's Horizon 2020 research and innovation programme grant agreement 101019620 (ERC Advanced Grant TOPUP). This work is also supported in part by the UK Science and Technology Facilities Council (STFC) through grant ST/T001011/1 and in part by the Deutsche Forschungsgemeinschaft (DFG, German Research Foundation) under grant 396021762-TRR 257. H. X. Z. is supported by the Natural Science Foundation of China (NSFC) under contract No. 11975200.
\end{acknowledgments}

\bibliography{WN3LO}

\begin{thebibliography}{69}%
\makeatletter
\providecommand \@ifxundefined [1]{%
 \@ifx{#1\undefined}
}%
\providecommand \@ifnum [1]{%
 \ifnum #1\expandafter \@firstoftwo
 \else \expandafter \@secondoftwo
 \fi
}%
\providecommand \@ifx [1]{%
 \ifx #1\expandafter \@firstoftwo
 \else \expandafter \@secondoftwo
 \fi
}%
\providecommand \natexlab [1]{#1}%
\providecommand \enquote  [1]{``#1''}%
\providecommand \bibnamefont  [1]{#1}%
\providecommand \bibfnamefont [1]{#1}%
\providecommand \citenamefont [1]{#1}%
\providecommand \href@noop [0]{\@secondoftwo}%
\providecommand \href [0]{\begingroup \@sanitize@url \@href}%
\providecommand \@href[1]{\@@startlink{#1}\@@href}%
\providecommand \@@href[1]{\endgroup#1\@@endlink}%
\providecommand \@sanitize@url [0]{\catcode `\\12\catcode `\$12\catcode
  `\&12\catcode `\#12\catcode `\^12\catcode `\_12\catcode `\%12\relax}%
\providecommand \@@startlink[1]{}%
\providecommand \@@endlink[0]{}%
\providecommand \url  [0]{\begingroup\@sanitize@url \@url }%
\providecommand \@url [1]{\endgroup\@href {#1}{\urlprefix }}%
\providecommand \urlprefix  [0]{URL }%
\providecommand \Eprint [0]{\href }%
\providecommand \doibase [0]{http://dx.doi.org/}%
\providecommand \selectlanguage [0]{\@gobble}%
\providecommand \bibinfo  [0]{\@secondoftwo}%
\providecommand \bibfield  [0]{\@secondoftwo}%
\providecommand \translation [1]{[#1]}%
\providecommand \BibitemOpen [0]{}%
\providecommand \bibitemStop [0]{}%
\providecommand \bibitemNoStop [0]{.\EOS\space}%
\providecommand \EOS [0]{\spacefactor3000\relax}%
\providecommand \BibitemShut  [1]{\csname bibitem#1\endcsname}%
\let\auto@bib@innerbib\@empty
\bibitem [{\citenamefont {Drell}\ and\ \citenamefont
  {Yan}(1970)}]{Drell:1970wh}%
  \BibitemOpen
  \bibfield  {author} {\bibinfo {author} {\bibfnamefont {S.~D.}\ \bibnamefont
  {Drell}}\ and\ \bibinfo {author} {\bibfnamefont {T.-M.}\ \bibnamefont
  {Yan}},\ }\href {\doibase 10.1103/PhysRevLett.25.316} {\bibfield  {journal}
  {\bibinfo  {journal} {Phys. Rev. Lett.}\ }\textbf {\bibinfo {volume} {25}},\
  \bibinfo {pages} {316} (\bibinfo {year} {1970})},\ \bibinfo {note} {[Erratum:
  Phys.Rev.Lett. 25, 902 (1970)]}\BibitemShut {NoStop}%
\bibitem [{\citenamefont {Aaltonen}\ \emph {et~al.}(2012)\citenamefont
  {Aaltonen} \emph {et~al.}}]{CDF:2012gpf}%
  \BibitemOpen
  \bibfield  {author} {\bibinfo {author} {\bibfnamefont {T.}~\bibnamefont
  {Aaltonen}} \emph {et~al.} (\bibinfo {collaboration} {CDF}),\ }\href
  {\doibase 10.1103/PhysRevLett.108.151803} {\bibfield  {journal} {\bibinfo
  {journal} {Phys. Rev. Lett.}\ }\textbf {\bibinfo {volume} {108}},\ \bibinfo
  {pages} {151803} (\bibinfo {year} {2012})},\ \Eprint
  {http://arxiv.org/abs/1203.0275} {arXiv:1203.0275 [hep-ex]} \BibitemShut
  {NoStop}%
\bibitem [{\citenamefont {Abazov}\ \emph {et~al.}(2012)\citenamefont {Abazov}
  \emph {et~al.}}]{D0:2012kms}%
  \BibitemOpen
  \bibfield  {author} {\bibinfo {author} {\bibfnamefont {V.~M.}\ \bibnamefont
  {Abazov}} \emph {et~al.} (\bibinfo {collaboration} {D0}),\ }\href {\doibase
  10.1103/PhysRevLett.108.151804} {\bibfield  {journal} {\bibinfo  {journal}
  {Phys. Rev. Lett.}\ }\textbf {\bibinfo {volume} {108}},\ \bibinfo {pages}
  {151804} (\bibinfo {year} {2012})},\ \Eprint {http://arxiv.org/abs/1203.0293}
  {arXiv:1203.0293 [hep-ex]} \BibitemShut {NoStop}%
\bibitem [{\citenamefont {Aaboud}\ \emph {et~al.}(2018)\citenamefont {Aaboud}
  \emph {et~al.}}]{ATLAS:2017rzl}%
  \BibitemOpen
  \bibfield  {author} {\bibinfo {author} {\bibfnamefont {M.}~\bibnamefont
  {Aaboud}} \emph {et~al.} (\bibinfo {collaboration} {ATLAS}),\ }\href
  {\doibase 10.1140/epjc/s10052-017-5475-4} {\bibfield  {journal} {\bibinfo
  {journal} {Eur. Phys. J. C}\ }\textbf {\bibinfo {volume} {78}},\ \bibinfo
  {pages} {110} (\bibinfo {year} {2018})},\ \bibinfo {note} {[Erratum:
  Eur.Phys.J.C 78, 898 (2018)]},\ \Eprint {http://arxiv.org/abs/1701.07240}
  {arXiv:1701.07240 [hep-ex]} \BibitemShut {NoStop}%
\bibitem [{\citenamefont {Aaij}\ \emph {et~al.}(2022)\citenamefont {Aaij} \emph
  {et~al.}}]{LHCb:2021bjt}%
  \BibitemOpen
  \bibfield  {author} {\bibinfo {author} {\bibfnamefont {R.}~\bibnamefont
  {Aaij}} \emph {et~al.} (\bibinfo {collaboration} {LHCb}),\ }\href {\doibase
  10.1007/JHEP01(2022)036} {\bibfield  {journal} {\bibinfo  {journal} {JHEP}\
  }\textbf {\bibinfo {volume} {01}},\ \bibinfo {pages} {036} (\bibinfo {year}
  {2022})},\ \Eprint {http://arxiv.org/abs/2109.01113} {arXiv:2109.01113
  [hep-ex]} \BibitemShut {NoStop}%
\bibitem [{\citenamefont {Aaltonen}\ \emph {et~al.}(2018)\citenamefont
  {Aaltonen} \emph {et~al.}}]{CDF:2018cnj}%
  \BibitemOpen
  \bibfield  {author} {\bibinfo {author} {\bibfnamefont {T.~A.}\ \bibnamefont
  {Aaltonen}} \emph {et~al.} (\bibinfo {collaboration} {CDF, D0}),\ }\href
  {\doibase 10.1103/PhysRevD.97.112007} {\bibfield  {journal} {\bibinfo
  {journal} {Phys. Rev. D}\ }\textbf {\bibinfo {volume} {97}},\ \bibinfo
  {pages} {112007} (\bibinfo {year} {2018})},\ \Eprint
  {http://arxiv.org/abs/1801.06283} {arXiv:1801.06283 [hep-ex]} \BibitemShut
  {NoStop}%
\bibitem [{\citenamefont {Aad}\ \emph {et~al.}(2019)\citenamefont {Aad} \emph
  {et~al.}}]{ATLAS:2019fgb}%
  \BibitemOpen
  \bibfield  {author} {\bibinfo {author} {\bibfnamefont {G.}~\bibnamefont
  {Aad}} \emph {et~al.} (\bibinfo {collaboration} {ATLAS}),\ }\href {\doibase
  10.1140/epjc/s10052-019-7199-0} {\bibfield  {journal} {\bibinfo  {journal}
  {Eur. Phys. J. C}\ }\textbf {\bibinfo {volume} {79}},\ \bibinfo {pages} {760}
  (\bibinfo {year} {2019})},\ \Eprint {http://arxiv.org/abs/1904.05631}
  {arXiv:1904.05631 [hep-ex]} \BibitemShut {NoStop}%
\bibitem [{\citenamefont {Aad}\ \emph {et~al.}(2020)\citenamefont {Aad} \emph
  {et~al.}}]{ATLAS:2019zci}%
  \BibitemOpen
  \bibfield  {author} {\bibinfo {author} {\bibfnamefont {G.}~\bibnamefont
  {Aad}} \emph {et~al.} (\bibinfo {collaboration} {ATLAS}),\ }\href {\doibase
  10.1140/epjc/s10052-020-8001-z} {\bibfield  {journal} {\bibinfo  {journal}
  {Eur. Phys. J. C}\ }\textbf {\bibinfo {volume} {80}},\ \bibinfo {pages} {616}
  (\bibinfo {year} {2020})},\ \Eprint {http://arxiv.org/abs/1912.02844}
  {arXiv:1912.02844 [hep-ex]} \BibitemShut {NoStop}%
\bibitem [{\citenamefont {Sirunyan}\ \emph {et~al.}(2020)\citenamefont
  {Sirunyan} \emph {et~al.}}]{CMS:2020cph}%
  \BibitemOpen
  \bibfield  {author} {\bibinfo {author} {\bibfnamefont {A.~M.}\ \bibnamefont
  {Sirunyan}} \emph {et~al.} (\bibinfo {collaboration} {CMS}),\ }\href
  {\doibase 10.1103/PhysRevD.102.092012} {\bibfield  {journal} {\bibinfo
  {journal} {Phys. Rev. D}\ }\textbf {\bibinfo {volume} {102}},\ \bibinfo
  {pages} {092012} (\bibinfo {year} {2020})},\ \Eprint
  {http://arxiv.org/abs/2008.04174} {arXiv:2008.04174 [hep-ex]} \BibitemShut
  {NoStop}%
\bibitem [{\citenamefont {Aaltonen}\ \emph {et~al.}(2022)\citenamefont
  {Aaltonen} \emph {et~al.}}]{CDF:2022hxs}%
  \BibitemOpen
  \bibfield  {author} {\bibinfo {author} {\bibfnamefont {T.}~\bibnamefont
  {Aaltonen}} \emph {et~al.} (\bibinfo {collaboration} {CDF}),\ }\href
  {\doibase 10.1126/science.abk1781} {\bibfield  {journal} {\bibinfo  {journal}
  {Science}\ }\textbf {\bibinfo {volume} {376}},\ \bibinfo {pages} {170}
  (\bibinfo {year} {2022})}\BibitemShut {NoStop}%
\bibitem [{\citenamefont {Altarelli}\ \emph {et~al.}(1979)\citenamefont
  {Altarelli}, \citenamefont {Ellis},\ and\ \citenamefont
  {Martinelli}}]{Altarelli:1979ub}%
  \BibitemOpen
  \bibfield  {author} {\bibinfo {author} {\bibfnamefont {G.}~\bibnamefont
  {Altarelli}}, \bibinfo {author} {\bibfnamefont {R.~K.}\ \bibnamefont
  {Ellis}}, \ and\ \bibinfo {author} {\bibfnamefont {G.}~\bibnamefont
  {Martinelli}},\ }\href {\doibase 10.1016/0550-3213(79)90116-0} {\bibfield
  {journal} {\bibinfo  {journal} {Nucl. Phys. B}\ }\textbf {\bibinfo {volume}
  {157}},\ \bibinfo {pages} {461} (\bibinfo {year} {1979})}\BibitemShut
  {NoStop}%
\bibitem [{\citenamefont {Hamberg}\ \emph {et~al.}(1991)\citenamefont
  {Hamberg}, \citenamefont {van Neerven},\ and\ \citenamefont
  {Matsuura}}]{Hamberg:1990np}%
  \BibitemOpen
  \bibfield  {author} {\bibinfo {author} {\bibfnamefont {R.}~\bibnamefont
  {Hamberg}}, \bibinfo {author} {\bibfnamefont {W.~L.}\ \bibnamefont {van
  Neerven}}, \ and\ \bibinfo {author} {\bibfnamefont {T.}~\bibnamefont
  {Matsuura}},\ }\href {\doibase 10.1016/0550-3213(91)90064-5} {\bibfield
  {journal} {\bibinfo  {journal} {Nucl. Phys. B}\ }\textbf {\bibinfo {volume}
  {359}},\ \bibinfo {pages} {343} (\bibinfo {year} {1991})},\ \bibinfo {note}
  {[Erratum: Nucl.Phys.B 644, 403--404 (2002)]}\BibitemShut {NoStop}%
\bibitem [{\citenamefont {Duhr}\ \emph
  {et~al.}(2020{\natexlab{a}})\citenamefont {Duhr}, \citenamefont {Dulat},\
  and\ \citenamefont {Mistlberger}}]{Duhr:2020sdp}%
  \BibitemOpen
  \bibfield  {author} {\bibinfo {author} {\bibfnamefont {C.}~\bibnamefont
  {Duhr}}, \bibinfo {author} {\bibfnamefont {F.}~\bibnamefont {Dulat}}, \ and\
  \bibinfo {author} {\bibfnamefont {B.}~\bibnamefont {Mistlberger}},\ }\href
  {\doibase 10.1007/JHEP11(2020)143} {\bibfield  {journal} {\bibinfo  {journal}
  {JHEP}\ }\textbf {\bibinfo {volume} {11}},\ \bibinfo {pages} {143} (\bibinfo
  {year} {2020}{\natexlab{a}})},\ \Eprint {http://arxiv.org/abs/2007.13313}
  {arXiv:2007.13313 [hep-ph]} \BibitemShut {NoStop}%
\bibitem [{\citenamefont {Anastasiou}\ \emph {et~al.}(2004)\citenamefont
  {Anastasiou}, \citenamefont {Dixon}, \citenamefont {Melnikov},\ and\
  \citenamefont {Petriello}}]{Anastasiou:2003ds}%
  \BibitemOpen
  \bibfield  {author} {\bibinfo {author} {\bibfnamefont {C.}~\bibnamefont
  {Anastasiou}}, \bibinfo {author} {\bibfnamefont {L.~J.}\ \bibnamefont
  {Dixon}}, \bibinfo {author} {\bibfnamefont {K.}~\bibnamefont {Melnikov}}, \
  and\ \bibinfo {author} {\bibfnamefont {F.}~\bibnamefont {Petriello}},\ }\href
  {\doibase 10.1103/PhysRevD.69.094008} {\bibfield  {journal} {\bibinfo
  {journal} {Phys. Rev. D}\ }\textbf {\bibinfo {volume} {69}},\ \bibinfo
  {pages} {094008} (\bibinfo {year} {2004})},\ \Eprint
  {http://arxiv.org/abs/hep-ph/0312266} {arXiv:hep-ph/0312266} \BibitemShut
  {NoStop}%
\bibitem [{\citenamefont {Melnikov}\ and\ \citenamefont
  {Petriello}(2006)}]{Melnikov:2006kv}%
  \BibitemOpen
  \bibfield  {author} {\bibinfo {author} {\bibfnamefont {K.}~\bibnamefont
  {Melnikov}}\ and\ \bibinfo {author} {\bibfnamefont {F.}~\bibnamefont
  {Petriello}},\ }\href {\doibase 10.1103/PhysRevD.74.114017} {\bibfield
  {journal} {\bibinfo  {journal} {Phys. Rev. D}\ }\textbf {\bibinfo {volume}
  {74}},\ \bibinfo {pages} {114017} (\bibinfo {year} {2006})},\ \Eprint
  {http://arxiv.org/abs/hep-ph/0609070} {arXiv:hep-ph/0609070} \BibitemShut
  {NoStop}%
\bibitem [{\citenamefont {Catani}\ \emph {et~al.}(2009)\citenamefont {Catani},
  \citenamefont {Cieri}, \citenamefont {Ferrera}, \citenamefont {de~Florian},\
  and\ \citenamefont {Grazzini}}]{Catani:2009sm}%
  \BibitemOpen
  \bibfield  {author} {\bibinfo {author} {\bibfnamefont {S.}~\bibnamefont
  {Catani}}, \bibinfo {author} {\bibfnamefont {L.}~\bibnamefont {Cieri}},
  \bibinfo {author} {\bibfnamefont {G.}~\bibnamefont {Ferrera}}, \bibinfo
  {author} {\bibfnamefont {D.}~\bibnamefont {de~Florian}}, \ and\ \bibinfo
  {author} {\bibfnamefont {M.}~\bibnamefont {Grazzini}},\ }\href {\doibase
  10.1103/PhysRevLett.103.082001} {\bibfield  {journal} {\bibinfo  {journal}
  {Phys. Rev. Lett.}\ }\textbf {\bibinfo {volume} {103}},\ \bibinfo {pages}
  {082001} (\bibinfo {year} {2009})},\ \Eprint {http://arxiv.org/abs/0903.2120}
  {arXiv:0903.2120 [hep-ph]} \BibitemShut {NoStop}%
\bibitem [{\citenamefont {Catani}\ \emph {et~al.}(2010)\citenamefont {Catani},
  \citenamefont {Ferrera},\ and\ \citenamefont {Grazzini}}]{Catani:2010en}%
  \BibitemOpen
  \bibfield  {author} {\bibinfo {author} {\bibfnamefont {S.}~\bibnamefont
  {Catani}}, \bibinfo {author} {\bibfnamefont {G.}~\bibnamefont {Ferrera}}, \
  and\ \bibinfo {author} {\bibfnamefont {M.}~\bibnamefont {Grazzini}},\ }\href
  {\doibase 10.1007/JHEP05(2010)006} {\bibfield  {journal} {\bibinfo  {journal}
  {JHEP}\ }\textbf {\bibinfo {volume} {05}},\ \bibinfo {pages} {006} (\bibinfo
  {year} {2010})},\ \Eprint {http://arxiv.org/abs/1002.3115} {arXiv:1002.3115
  [hep-ph]} \BibitemShut {NoStop}%
\bibitem [{\citenamefont {Dittmaier}\ and\ \citenamefont
  {Kr\"amer}(2002)}]{Dittmaier:2001ay}%
  \BibitemOpen
  \bibfield  {author} {\bibinfo {author} {\bibfnamefont {S.}~\bibnamefont
  {Dittmaier}}\ and\ \bibinfo {author} {\bibfnamefont {M.}~\bibnamefont
  {Kr\"amer}},\ }\href {\doibase 10.1103/PhysRevD.65.073007} {\bibfield
  {journal} {\bibinfo  {journal} {Phys. Rev. D}\ }\textbf {\bibinfo {volume}
  {65}},\ \bibinfo {pages} {073007} (\bibinfo {year} {2002})},\ \Eprint
  {http://arxiv.org/abs/hep-ph/0109062} {arXiv:hep-ph/0109062} \BibitemShut
  {NoStop}%
\bibitem [{\citenamefont {Baur}\ and\ \citenamefont
  {Wackeroth}(2004)}]{Baur:2004ig}%
  \BibitemOpen
  \bibfield  {author} {\bibinfo {author} {\bibfnamefont {U.}~\bibnamefont
  {Baur}}\ and\ \bibinfo {author} {\bibfnamefont {D.}~\bibnamefont
  {Wackeroth}},\ }\href {\doibase 10.1103/PhysRevD.70.073015} {\bibfield
  {journal} {\bibinfo  {journal} {Phys. Rev. D}\ }\textbf {\bibinfo {volume}
  {70}},\ \bibinfo {pages} {073015} (\bibinfo {year} {2004})},\ \Eprint
  {http://arxiv.org/abs/hep-ph/0405191} {arXiv:hep-ph/0405191} \BibitemShut
  {NoStop}%
\bibitem [{\citenamefont {Carloni~Calame}\ \emph {et~al.}(2007)\citenamefont
  {Carloni~Calame}, \citenamefont {Montagna}, \citenamefont {Nicrosini},\ and\
  \citenamefont {Vicini}}]{CarloniCalame:2007cd}%
  \BibitemOpen
  \bibfield  {author} {\bibinfo {author} {\bibfnamefont {C.~M.}\ \bibnamefont
  {Carloni~Calame}}, \bibinfo {author} {\bibfnamefont {G.}~\bibnamefont
  {Montagna}}, \bibinfo {author} {\bibfnamefont {O.}~\bibnamefont {Nicrosini}},
  \ and\ \bibinfo {author} {\bibfnamefont {A.}~\bibnamefont {Vicini}},\ }\href
  {\doibase 10.1088/1126-6708/2007/10/109} {\bibfield  {journal} {\bibinfo
  {journal} {JHEP}\ }\textbf {\bibinfo {volume} {10}},\ \bibinfo {pages} {109}
  (\bibinfo {year} {2007})},\ \Eprint {http://arxiv.org/abs/0710.1722}
  {arXiv:0710.1722 [hep-ph]} \BibitemShut {NoStop}%
\bibitem [{\citenamefont {Dittmaier}\ \emph {et~al.}(2016)\citenamefont
  {Dittmaier}, \citenamefont {Huss},\ and\ \citenamefont
  {Schwinn}}]{Dittmaier:2015rxo}%
  \BibitemOpen
  \bibfield  {author} {\bibinfo {author} {\bibfnamefont {S.}~\bibnamefont
  {Dittmaier}}, \bibinfo {author} {\bibfnamefont {A.}~\bibnamefont {Huss}}, \
  and\ \bibinfo {author} {\bibfnamefont {C.}~\bibnamefont {Schwinn}},\ }\href
  {\doibase 10.1016/j.nuclphysb.2016.01.006} {\bibfield  {journal} {\bibinfo
  {journal} {Nucl. Phys. B}\ }\textbf {\bibinfo {volume} {904}},\ \bibinfo
  {pages} {216} (\bibinfo {year} {2016})},\ \Eprint
  {http://arxiv.org/abs/1511.08016} {arXiv:1511.08016 [hep-ph]} \BibitemShut
  {NoStop}%
\bibitem [{\citenamefont {Dittmaier}\ \emph {et~al.}(2020)\citenamefont
  {Dittmaier}, \citenamefont {Schmidt},\ and\ \citenamefont
  {Schwarz}}]{Dittmaier:2020vra}%
  \BibitemOpen
  \bibfield  {author} {\bibinfo {author} {\bibfnamefont {S.}~\bibnamefont
  {Dittmaier}}, \bibinfo {author} {\bibfnamefont {T.}~\bibnamefont {Schmidt}},
  \ and\ \bibinfo {author} {\bibfnamefont {J.}~\bibnamefont {Schwarz}},\ }\href
  {\doibase 10.1007/JHEP12(2020)201} {\bibfield  {journal} {\bibinfo  {journal}
  {JHEP}\ }\textbf {\bibinfo {volume} {12}},\ \bibinfo {pages} {201} (\bibinfo
  {year} {2020})},\ \Eprint {http://arxiv.org/abs/2009.02229} {arXiv:2009.02229
  [hep-ph]} \BibitemShut {NoStop}%
\bibitem [{\citenamefont {Behring}\ \emph
  {et~al.}(2021{\natexlab{a}})\citenamefont {Behring}, \citenamefont
  {Buccioni}, \citenamefont {Caola}, \citenamefont {Delto}, \citenamefont
  {Jaquier}, \citenamefont {Melnikov},\ and\ \citenamefont
  {R\"ontsch}}]{Behring:2020cqi}%
  \BibitemOpen
  \bibfield  {author} {\bibinfo {author} {\bibfnamefont {A.}~\bibnamefont
  {Behring}}, \bibinfo {author} {\bibfnamefont {F.}~\bibnamefont {Buccioni}},
  \bibinfo {author} {\bibfnamefont {F.}~\bibnamefont {Caola}}, \bibinfo
  {author} {\bibfnamefont {M.}~\bibnamefont {Delto}}, \bibinfo {author}
  {\bibfnamefont {M.}~\bibnamefont {Jaquier}}, \bibinfo {author} {\bibfnamefont
  {K.}~\bibnamefont {Melnikov}}, \ and\ \bibinfo {author} {\bibfnamefont
  {R.}~\bibnamefont {R\"ontsch}},\ }\href {\doibase
  10.1103/PhysRevD.103.013008} {\bibfield  {journal} {\bibinfo  {journal}
  {Phys. Rev. D}\ }\textbf {\bibinfo {volume} {103}},\ \bibinfo {pages}
  {013008} (\bibinfo {year} {2021}{\natexlab{a}})},\ \Eprint
  {http://arxiv.org/abs/2009.10386} {arXiv:2009.10386 [hep-ph]} \BibitemShut
  {NoStop}%
\bibitem [{\citenamefont {Buonocore}\ \emph {et~al.}(2021)\citenamefont
  {Buonocore}, \citenamefont {Grazzini}, \citenamefont {Kallweit},
  \citenamefont {Savoini},\ and\ \citenamefont
  {Tramontano}}]{Buonocore:2021rxx}%
  \BibitemOpen
  \bibfield  {author} {\bibinfo {author} {\bibfnamefont {L.}~\bibnamefont
  {Buonocore}}, \bibinfo {author} {\bibfnamefont {M.}~\bibnamefont {Grazzini}},
  \bibinfo {author} {\bibfnamefont {S.}~\bibnamefont {Kallweit}}, \bibinfo
  {author} {\bibfnamefont {C.}~\bibnamefont {Savoini}}, \ and\ \bibinfo
  {author} {\bibfnamefont {F.}~\bibnamefont {Tramontano}},\ }\href {\doibase
  10.1103/PhysRevD.103.114012} {\bibfield  {journal} {\bibinfo  {journal}
  {Phys. Rev. D}\ }\textbf {\bibinfo {volume} {103}},\ \bibinfo {pages}
  {114012} (\bibinfo {year} {2021})},\ \Eprint
  {http://arxiv.org/abs/2102.12539} {arXiv:2102.12539 [hep-ph]} \BibitemShut
  {NoStop}%
\bibitem [{\citenamefont {Behring}\ \emph
  {et~al.}(2021{\natexlab{b}})\citenamefont {Behring}, \citenamefont
  {Buccioni}, \citenamefont {Caola}, \citenamefont {Delto}, \citenamefont
  {Jaquier}, \citenamefont {Melnikov},\ and\ \citenamefont
  {R\"ontsch}}]{Behring:2021adr}%
  \BibitemOpen
  \bibfield  {author} {\bibinfo {author} {\bibfnamefont {A.}~\bibnamefont
  {Behring}}, \bibinfo {author} {\bibfnamefont {F.}~\bibnamefont {Buccioni}},
  \bibinfo {author} {\bibfnamefont {F.}~\bibnamefont {Caola}}, \bibinfo
  {author} {\bibfnamefont {M.}~\bibnamefont {Delto}}, \bibinfo {author}
  {\bibfnamefont {M.}~\bibnamefont {Jaquier}}, \bibinfo {author} {\bibfnamefont
  {K.}~\bibnamefont {Melnikov}}, \ and\ \bibinfo {author} {\bibfnamefont
  {R.}~\bibnamefont {R\"ontsch}},\ }\href {\doibase
  10.1103/PhysRevD.103.113002} {\bibfield  {journal} {\bibinfo  {journal}
  {Phys. Rev. D}\ }\textbf {\bibinfo {volume} {103}},\ \bibinfo {pages}
  {113002} (\bibinfo {year} {2021}{\natexlab{b}})},\ \Eprint
  {http://arxiv.org/abs/2103.02671} {arXiv:2103.02671 [hep-ph]} \BibitemShut
  {NoStop}%
\bibitem [{\citenamefont {Balazs}\ and\ \citenamefont
  {Yuan}(1997)}]{Balazs:1997xd}%
  \BibitemOpen
  \bibfield  {author} {\bibinfo {author} {\bibfnamefont {C.}~\bibnamefont
  {Balazs}}\ and\ \bibinfo {author} {\bibfnamefont {C.~P.}\ \bibnamefont
  {Yuan}},\ }\href {\doibase 10.1103/PhysRevD.56.5558} {\bibfield  {journal}
  {\bibinfo  {journal} {Phys. Rev. D}\ }\textbf {\bibinfo {volume} {56}},\
  \bibinfo {pages} {5558} (\bibinfo {year} {1997})},\ \Eprint
  {http://arxiv.org/abs/hep-ph/9704258} {arXiv:hep-ph/9704258} \BibitemShut
  {NoStop}%
\bibitem [{\citenamefont {Bozzi}\ \emph {et~al.}(2011)\citenamefont {Bozzi},
  \citenamefont {Catani}, \citenamefont {Ferrera}, \citenamefont {de~Florian},\
  and\ \citenamefont {Grazzini}}]{Bozzi:2010xn}%
  \BibitemOpen
  \bibfield  {author} {\bibinfo {author} {\bibfnamefont {G.}~\bibnamefont
  {Bozzi}}, \bibinfo {author} {\bibfnamefont {S.}~\bibnamefont {Catani}},
  \bibinfo {author} {\bibfnamefont {G.}~\bibnamefont {Ferrera}}, \bibinfo
  {author} {\bibfnamefont {D.}~\bibnamefont {de~Florian}}, \ and\ \bibinfo
  {author} {\bibfnamefont {M.}~\bibnamefont {Grazzini}},\ }\href {\doibase
  10.1016/j.physletb.2010.12.024} {\bibfield  {journal} {\bibinfo  {journal}
  {Phys. Lett. B}\ }\textbf {\bibinfo {volume} {696}},\ \bibinfo {pages} {207}
  (\bibinfo {year} {2011})},\ \Eprint {http://arxiv.org/abs/1007.2351}
  {arXiv:1007.2351 [hep-ph]} \BibitemShut {NoStop}%
\bibitem [{\citenamefont {Becher}\ \emph {et~al.}(2012)\citenamefont {Becher},
  \citenamefont {Neubert},\ and\ \citenamefont {Wilhelm}}]{Becher:2011xn}%
  \BibitemOpen
  \bibfield  {author} {\bibinfo {author} {\bibfnamefont {T.}~\bibnamefont
  {Becher}}, \bibinfo {author} {\bibfnamefont {M.}~\bibnamefont {Neubert}}, \
  and\ \bibinfo {author} {\bibfnamefont {D.}~\bibnamefont {Wilhelm}},\ }\href
  {\doibase 10.1007/JHEP02(2012)124} {\bibfield  {journal} {\bibinfo  {journal}
  {JHEP}\ }\textbf {\bibinfo {volume} {02}},\ \bibinfo {pages} {124} (\bibinfo
  {year} {2012})},\ \Eprint {http://arxiv.org/abs/1109.6027} {arXiv:1109.6027
  [hep-ph]} \BibitemShut {NoStop}%
\bibitem [{\citenamefont {Bizon}\ \emph {et~al.}(2019)\citenamefont {Bizon},
  \citenamefont {Gehrmann-De~Ridder}, \citenamefont {Gehrmann}, \citenamefont
  {Glover}, \citenamefont {Huss}, \citenamefont {Monni}, \citenamefont {Re},
  \citenamefont {Rottoli},\ and\ \citenamefont {Walker}}]{Bizon:2019zgf}%
  \BibitemOpen
  \bibfield  {author} {\bibinfo {author} {\bibfnamefont {W.}~\bibnamefont
  {Bizon}}, \bibinfo {author} {\bibfnamefont {A.}~\bibnamefont
  {Gehrmann-De~Ridder}}, \bibinfo {author} {\bibfnamefont {T.}~\bibnamefont
  {Gehrmann}}, \bibinfo {author} {\bibfnamefont {N.}~\bibnamefont {Glover}},
  \bibinfo {author} {\bibfnamefont {A.}~\bibnamefont {Huss}}, \bibinfo {author}
  {\bibfnamefont {P.~F.}\ \bibnamefont {Monni}}, \bibinfo {author}
  {\bibfnamefont {E.}~\bibnamefont {Re}}, \bibinfo {author} {\bibfnamefont
  {L.}~\bibnamefont {Rottoli}}, \ and\ \bibinfo {author} {\bibfnamefont
  {D.~M.}\ \bibnamefont {Walker}},\ }\href {\doibase
  10.1140/epjc/s10052-019-7324-0} {\bibfield  {journal} {\bibinfo  {journal}
  {Eur. Phys. J. C}\ }\textbf {\bibinfo {volume} {79}},\ \bibinfo {pages} {868}
  (\bibinfo {year} {2019})},\ \Eprint {http://arxiv.org/abs/1905.05171}
  {arXiv:1905.05171 [hep-ph]} \BibitemShut {NoStop}%
\bibitem [{\citenamefont {Catani}\ and\ \citenamefont
  {Grazzini}(2007)}]{Catani:2007vq}%
  \BibitemOpen
  \bibfield  {author} {\bibinfo {author} {\bibfnamefont {S.}~\bibnamefont
  {Catani}}\ and\ \bibinfo {author} {\bibfnamefont {M.}~\bibnamefont
  {Grazzini}},\ }\href {\doibase 10.1103/PhysRevLett.98.222002} {\bibfield
  {journal} {\bibinfo  {journal} {Phys. Rev. Lett.}\ }\textbf {\bibinfo
  {volume} {98}},\ \bibinfo {pages} {222002} (\bibinfo {year} {2007})},\
  \Eprint {http://arxiv.org/abs/hep-ph/0703012} {arXiv:hep-ph/0703012}
  \BibitemShut {NoStop}%
\bibitem [{\citenamefont {Cieri}\ \emph {et~al.}(2019)\citenamefont {Cieri},
  \citenamefont {Chen}, \citenamefont {Gehrmann}, \citenamefont {Glover},\ and\
  \citenamefont {Huss}}]{Cieri:2018oms}%
  \BibitemOpen
  \bibfield  {author} {\bibinfo {author} {\bibfnamefont {L.}~\bibnamefont
  {Cieri}}, \bibinfo {author} {\bibfnamefont {X.}~\bibnamefont {Chen}},
  \bibinfo {author} {\bibfnamefont {T.}~\bibnamefont {Gehrmann}}, \bibinfo
  {author} {\bibfnamefont {E.~W.~N.}\ \bibnamefont {Glover}}, \ and\ \bibinfo
  {author} {\bibfnamefont {A.}~\bibnamefont {Huss}},\ }\href {\doibase
  10.1007/JHEP02(2019)096} {\bibfield  {journal} {\bibinfo  {journal} {JHEP}\
  }\textbf {\bibinfo {volume} {02}},\ \bibinfo {pages} {096} (\bibinfo {year}
  {2019})},\ \Eprint {http://arxiv.org/abs/1807.11501} {arXiv:1807.11501
  [hep-ph]} \BibitemShut {NoStop}%
\bibitem [{\citenamefont {Billis}\ \emph {et~al.}(2021)\citenamefont {Billis},
  \citenamefont {Ebert}, \citenamefont {Michel},\ and\ \citenamefont
  {Tackmann}}]{Billis:2019vxg}%
  \BibitemOpen
  \bibfield  {author} {\bibinfo {author} {\bibfnamefont {G.}~\bibnamefont
  {Billis}}, \bibinfo {author} {\bibfnamefont {M.~A.}\ \bibnamefont {Ebert}},
  \bibinfo {author} {\bibfnamefont {J.~K.~L.}\ \bibnamefont {Michel}}, \ and\
  \bibinfo {author} {\bibfnamefont {F.~J.}\ \bibnamefont {Tackmann}},\ }\href
  {\doibase 10.1140/epjp/s13360-021-01155-y} {\bibfield  {journal} {\bibinfo
  {journal} {Eur. Phys. J. Plus}\ }\textbf {\bibinfo {volume} {136}},\ \bibinfo
  {pages} {214} (\bibinfo {year} {2021})},\ \Eprint
  {http://arxiv.org/abs/1909.00811} {arXiv:1909.00811 [hep-ph]} \BibitemShut
  {NoStop}%
\bibitem [{\citenamefont {Chen}\ \emph
  {et~al.}(2022{\natexlab{a}})\citenamefont {Chen}, \citenamefont {Gehrmann},
  \citenamefont {Glover}, \citenamefont {Huss}, \citenamefont {Yang},\ and\
  \citenamefont {Zhu}}]{Chen:2021vtu}%
  \BibitemOpen
  \bibfield  {author} {\bibinfo {author} {\bibfnamefont {X.}~\bibnamefont
  {Chen}}, \bibinfo {author} {\bibfnamefont {T.}~\bibnamefont {Gehrmann}},
  \bibinfo {author} {\bibfnamefont {N.}~\bibnamefont {Glover}}, \bibinfo
  {author} {\bibfnamefont {A.}~\bibnamefont {Huss}}, \bibinfo {author}
  {\bibfnamefont {T.-Z.}\ \bibnamefont {Yang}}, \ and\ \bibinfo {author}
  {\bibfnamefont {H.~X.}\ \bibnamefont {Zhu}},\ }\href {\doibase
  10.1103/PhysRevLett.128.052001} {\bibfield  {journal} {\bibinfo  {journal}
  {Phys. Rev. Lett.}\ }\textbf {\bibinfo {volume} {128}},\ \bibinfo {pages}
  {052001} (\bibinfo {year} {2022}{\natexlab{a}})},\ \Eprint
  {http://arxiv.org/abs/2107.09085} {arXiv:2107.09085 [hep-ph]} \BibitemShut
  {NoStop}%
\bibitem [{\citenamefont {Bauer}\ \emph {et~al.}(2000)\citenamefont {Bauer},
  \citenamefont {Fleming},\ and\ \citenamefont {Luke}}]{Bauer:2000ew}%
  \BibitemOpen
  \bibfield  {author} {\bibinfo {author} {\bibfnamefont {C.~W.}\ \bibnamefont
  {Bauer}}, \bibinfo {author} {\bibfnamefont {S.}~\bibnamefont {Fleming}}, \
  and\ \bibinfo {author} {\bibfnamefont {M.~E.}\ \bibnamefont {Luke}},\ }\href
  {\doibase 10.1103/PhysRevD.63.014006} {\bibfield  {journal} {\bibinfo
  {journal} {Phys. Rev. D}\ }\textbf {\bibinfo {volume} {63}},\ \bibinfo
  {pages} {014006} (\bibinfo {year} {2000})},\ \Eprint
  {http://arxiv.org/abs/hep-ph/0005275} {arXiv:hep-ph/0005275} \BibitemShut
  {NoStop}%
\bibitem [{\citenamefont {Bauer}\ \emph {et~al.}(2001)\citenamefont {Bauer},
  \citenamefont {Fleming}, \citenamefont {Pirjol},\ and\ \citenamefont
  {Stewart}}]{Bauer:2000yr}%
  \BibitemOpen
  \bibfield  {author} {\bibinfo {author} {\bibfnamefont {C.~W.}\ \bibnamefont
  {Bauer}}, \bibinfo {author} {\bibfnamefont {S.}~\bibnamefont {Fleming}},
  \bibinfo {author} {\bibfnamefont {D.}~\bibnamefont {Pirjol}}, \ and\ \bibinfo
  {author} {\bibfnamefont {I.~W.}\ \bibnamefont {Stewart}},\ }\href {\doibase
  10.1103/PhysRevD.63.114020} {\bibfield  {journal} {\bibinfo  {journal} {Phys.
  Rev. D}\ }\textbf {\bibinfo {volume} {63}},\ \bibinfo {pages} {114020}
  (\bibinfo {year} {2001})},\ \Eprint {http://arxiv.org/abs/hep-ph/0011336}
  {arXiv:hep-ph/0011336} \BibitemShut {NoStop}%
\bibitem [{\citenamefont {Bauer}\ \emph
  {et~al.}(2002{\natexlab{a}})\citenamefont {Bauer}, \citenamefont {Pirjol},\
  and\ \citenamefont {Stewart}}]{Bauer:2001yt}%
  \BibitemOpen
  \bibfield  {author} {\bibinfo {author} {\bibfnamefont {C.~W.}\ \bibnamefont
  {Bauer}}, \bibinfo {author} {\bibfnamefont {D.}~\bibnamefont {Pirjol}}, \
  and\ \bibinfo {author} {\bibfnamefont {I.~W.}\ \bibnamefont {Stewart}},\
  }\href {\doibase 10.1103/PhysRevD.65.054022} {\bibfield  {journal} {\bibinfo
  {journal} {Phys. Rev. D}\ }\textbf {\bibinfo {volume} {65}},\ \bibinfo
  {pages} {054022} (\bibinfo {year} {2002}{\natexlab{a}})},\ \Eprint
  {http://arxiv.org/abs/hep-ph/0109045} {arXiv:hep-ph/0109045} \BibitemShut
  {NoStop}%
\bibitem [{\citenamefont {Bauer}\ \emph
  {et~al.}(2002{\natexlab{b}})\citenamefont {Bauer}, \citenamefont {Fleming},
  \citenamefont {Pirjol}, \citenamefont {Rothstein},\ and\ \citenamefont
  {Stewart}}]{Bauer:2002nz}%
  \BibitemOpen
  \bibfield  {author} {\bibinfo {author} {\bibfnamefont {C.~W.}\ \bibnamefont
  {Bauer}}, \bibinfo {author} {\bibfnamefont {S.}~\bibnamefont {Fleming}},
  \bibinfo {author} {\bibfnamefont {D.}~\bibnamefont {Pirjol}}, \bibinfo
  {author} {\bibfnamefont {I.~Z.}\ \bibnamefont {Rothstein}}, \ and\ \bibinfo
  {author} {\bibfnamefont {I.~W.}\ \bibnamefont {Stewart}},\ }\href {\doibase
  10.1103/PhysRevD.66.014017} {\bibfield  {journal} {\bibinfo  {journal} {Phys.
  Rev. D}\ }\textbf {\bibinfo {volume} {66}},\ \bibinfo {pages} {014017}
  (\bibinfo {year} {2002}{\natexlab{b}})},\ \Eprint
  {http://arxiv.org/abs/hep-ph/0202088} {arXiv:hep-ph/0202088} \BibitemShut
  {NoStop}%
\bibitem [{\citenamefont {Beneke}\ \emph {et~al.}(2002)\citenamefont {Beneke},
  \citenamefont {Chapovsky}, \citenamefont {Diehl},\ and\ \citenamefont
  {Feldmann}}]{Beneke:2002ph}%
  \BibitemOpen
  \bibfield  {author} {\bibinfo {author} {\bibfnamefont {M.}~\bibnamefont
  {Beneke}}, \bibinfo {author} {\bibfnamefont {A.~P.}\ \bibnamefont
  {Chapovsky}}, \bibinfo {author} {\bibfnamefont {M.}~\bibnamefont {Diehl}}, \
  and\ \bibinfo {author} {\bibfnamefont {T.}~\bibnamefont {Feldmann}},\ }\href
  {\doibase 10.1016/S0550-3213(02)00687-9} {\bibfield  {journal} {\bibinfo
  {journal} {Nucl. Phys. B}\ }\textbf {\bibinfo {volume} {643}},\ \bibinfo
  {pages} {431} (\bibinfo {year} {2002})},\ \Eprint
  {http://arxiv.org/abs/hep-ph/0206152} {arXiv:hep-ph/0206152} \BibitemShut
  {NoStop}%
\bibitem [{\citenamefont {Chen}\ \emph {et~al.}(2019)\citenamefont {Chen},
  \citenamefont {Gehrmann}, \citenamefont {Glover}, \citenamefont {Huss},
  \citenamefont {Li}, \citenamefont {Neill}, \citenamefont {Schulze},
  \citenamefont {Stewart},\ and\ \citenamefont {Zhu}}]{Chen:2018pzu}%
  \BibitemOpen
  \bibfield  {author} {\bibinfo {author} {\bibfnamefont {X.}~\bibnamefont
  {Chen}}, \bibinfo {author} {\bibfnamefont {T.}~\bibnamefont {Gehrmann}},
  \bibinfo {author} {\bibfnamefont {E.~W.~N.}\ \bibnamefont {Glover}}, \bibinfo
  {author} {\bibfnamefont {A.}~\bibnamefont {Huss}}, \bibinfo {author}
  {\bibfnamefont {Y.}~\bibnamefont {Li}}, \bibinfo {author} {\bibfnamefont
  {D.}~\bibnamefont {Neill}}, \bibinfo {author} {\bibfnamefont
  {M.}~\bibnamefont {Schulze}}, \bibinfo {author} {\bibfnamefont {I.~W.}\
  \bibnamefont {Stewart}}, \ and\ \bibinfo {author} {\bibfnamefont {H.~X.}\
  \bibnamefont {Zhu}},\ }\href {\doibase 10.1016/j.physletb.2018.11.037}
  {\bibfield  {journal} {\bibinfo  {journal} {Phys. Lett. B}\ }\textbf
  {\bibinfo {volume} {788}},\ \bibinfo {pages} {425} (\bibinfo {year}
  {2019})},\ \Eprint {http://arxiv.org/abs/1805.00736} {arXiv:1805.00736
  [hep-ph]} \BibitemShut {NoStop}%
\bibitem [{\citenamefont {Chiu}\ \emph {et~al.}(2012)\citenamefont {Chiu},
  \citenamefont {Jain}, \citenamefont {Neill},\ and\ \citenamefont
  {Rothstein}}]{Chiu:2012ir}%
  \BibitemOpen
  \bibfield  {author} {\bibinfo {author} {\bibfnamefont {J.-Y.}\ \bibnamefont
  {Chiu}}, \bibinfo {author} {\bibfnamefont {A.}~\bibnamefont {Jain}}, \bibinfo
  {author} {\bibfnamefont {D.}~\bibnamefont {Neill}}, \ and\ \bibinfo {author}
  {\bibfnamefont {I.~Z.}\ \bibnamefont {Rothstein}},\ }\href {\doibase
  10.1007/JHEP05(2012)084} {\bibfield  {journal} {\bibinfo  {journal} {JHEP}\
  }\textbf {\bibinfo {volume} {05}},\ \bibinfo {pages} {084} (\bibinfo {year}
  {2012})},\ \Eprint {http://arxiv.org/abs/1202.0814} {arXiv:1202.0814
  [hep-ph]} \BibitemShut {NoStop}%
\bibitem [{\citenamefont {Li}\ and\ \citenamefont {Zhu}(2017)}]{Li:2016ctv}%
  \BibitemOpen
  \bibfield  {author} {\bibinfo {author} {\bibfnamefont {Y.}~\bibnamefont
  {Li}}\ and\ \bibinfo {author} {\bibfnamefont {H.~X.}\ \bibnamefont {Zhu}},\
  }\href {\doibase 10.1103/PhysRevLett.118.022004} {\bibfield  {journal}
  {\bibinfo  {journal} {Phys. Rev. Lett.}\ }\textbf {\bibinfo {volume} {118}},\
  \bibinfo {pages} {022004} (\bibinfo {year} {2017})},\ \Eprint
  {http://arxiv.org/abs/1604.01404} {arXiv:1604.01404 [hep-ph]} \BibitemShut
  {NoStop}%
\bibitem [{\citenamefont {Luo}\ \emph {et~al.}(2020)\citenamefont {Luo},
  \citenamefont {Yang}, \citenamefont {Zhu},\ and\ \citenamefont
  {Zhu}}]{Luo:2019szz}%
  \BibitemOpen
  \bibfield  {author} {\bibinfo {author} {\bibfnamefont {M.-x.}\ \bibnamefont
  {Luo}}, \bibinfo {author} {\bibfnamefont {T.-Z.}\ \bibnamefont {Yang}},
  \bibinfo {author} {\bibfnamefont {H.~X.}\ \bibnamefont {Zhu}}, \ and\
  \bibinfo {author} {\bibfnamefont {Y.~J.}\ \bibnamefont {Zhu}},\ }\href
  {\doibase 10.1103/PhysRevLett.124.092001} {\bibfield  {journal} {\bibinfo
  {journal} {Phys. Rev. Lett.}\ }\textbf {\bibinfo {volume} {124}},\ \bibinfo
  {pages} {092001} (\bibinfo {year} {2020})},\ \Eprint
  {http://arxiv.org/abs/1912.05778} {arXiv:1912.05778 [hep-ph]} \BibitemShut
  {NoStop}%
\bibitem [{\citenamefont {Ebert}\ \emph {et~al.}(2020)\citenamefont {Ebert},
  \citenamefont {Mistlberger},\ and\ \citenamefont {Vita}}]{Ebert:2020yqt}%
  \BibitemOpen
  \bibfield  {author} {\bibinfo {author} {\bibfnamefont {M.~A.}\ \bibnamefont
  {Ebert}}, \bibinfo {author} {\bibfnamefont {B.}~\bibnamefont {Mistlberger}},
  \ and\ \bibinfo {author} {\bibfnamefont {G.}~\bibnamefont {Vita}},\ }\href
  {\doibase 10.1007/JHEP09(2020)146} {\bibfield  {journal} {\bibinfo  {journal}
  {JHEP}\ }\textbf {\bibinfo {volume} {09}},\ \bibinfo {pages} {146} (\bibinfo
  {year} {2020})},\ \Eprint {http://arxiv.org/abs/2006.05329} {arXiv:2006.05329
  [hep-ph]} \BibitemShut {NoStop}%
\bibitem [{\citenamefont {Luo}\ \emph {et~al.}(2021)\citenamefont {Luo},
  \citenamefont {Yang}, \citenamefont {Zhu},\ and\ \citenamefont
  {Zhu}}]{Luo:2020epw}%
  \BibitemOpen
  \bibfield  {author} {\bibinfo {author} {\bibfnamefont {M.-x.}\ \bibnamefont
  {Luo}}, \bibinfo {author} {\bibfnamefont {T.-Z.}\ \bibnamefont {Yang}},
  \bibinfo {author} {\bibfnamefont {H.~X.}\ \bibnamefont {Zhu}}, \ and\
  \bibinfo {author} {\bibfnamefont {Y.~J.}\ \bibnamefont {Zhu}},\ }\href
  {\doibase 10.1007/JHEP06(2021)115} {\bibfield  {journal} {\bibinfo  {journal}
  {JHEP}\ }\textbf {\bibinfo {volume} {06}},\ \bibinfo {pages} {115} (\bibinfo
  {year} {2021})},\ \Eprint {http://arxiv.org/abs/2012.03256} {arXiv:2012.03256
  [hep-ph]} \BibitemShut {NoStop}%
\bibitem [{\citenamefont {Baikov}\ \emph {et~al.}(2009)\citenamefont {Baikov},
  \citenamefont {Chetyrkin}, \citenamefont {Smirnov}, \citenamefont {Smirnov},\
  and\ \citenamefont {Steinhauser}}]{Baikov:2009bg}%
  \BibitemOpen
  \bibfield  {author} {\bibinfo {author} {\bibfnamefont {P.~A.}\ \bibnamefont
  {Baikov}}, \bibinfo {author} {\bibfnamefont {K.~G.}\ \bibnamefont
  {Chetyrkin}}, \bibinfo {author} {\bibfnamefont {A.~V.}\ \bibnamefont
  {Smirnov}}, \bibinfo {author} {\bibfnamefont {V.~A.}\ \bibnamefont
  {Smirnov}}, \ and\ \bibinfo {author} {\bibfnamefont {M.}~\bibnamefont
  {Steinhauser}},\ }\href {\doibase 10.1103/PhysRevLett.102.212002} {\bibfield
  {journal} {\bibinfo  {journal} {Phys. Rev. Lett.}\ }\textbf {\bibinfo
  {volume} {102}},\ \bibinfo {pages} {212002} (\bibinfo {year} {2009})},\
  \Eprint {http://arxiv.org/abs/0902.3519} {arXiv:0902.3519 [hep-ph]}
  \BibitemShut {NoStop}%
\bibitem [{\citenamefont {Lee}\ \emph {et~al.}(2010)\citenamefont {Lee},
  \citenamefont {Smirnov},\ and\ \citenamefont {Smirnov}}]{Lee:2010cga}%
  \BibitemOpen
  \bibfield  {author} {\bibinfo {author} {\bibfnamefont {R.~N.}\ \bibnamefont
  {Lee}}, \bibinfo {author} {\bibfnamefont {A.~V.}\ \bibnamefont {Smirnov}}, \
  and\ \bibinfo {author} {\bibfnamefont {V.~A.}\ \bibnamefont {Smirnov}},\
  }\href {\doibase 10.1007/JHEP04(2010)020} {\bibfield  {journal} {\bibinfo
  {journal} {JHEP}\ }\textbf {\bibinfo {volume} {04}},\ \bibinfo {pages} {020}
  (\bibinfo {year} {2010})},\ \Eprint {http://arxiv.org/abs/1001.2887}
  {arXiv:1001.2887 [hep-ph]} \BibitemShut {NoStop}%
\bibitem [{\citenamefont {Gehrmann}\ \emph {et~al.}(2010)\citenamefont
  {Gehrmann}, \citenamefont {Glover}, \citenamefont {Huber}, \citenamefont
  {Ikizlerli},\ and\ \citenamefont {Studerus}}]{Gehrmann:2010ue}%
  \BibitemOpen
  \bibfield  {author} {\bibinfo {author} {\bibfnamefont {T.}~\bibnamefont
  {Gehrmann}}, \bibinfo {author} {\bibfnamefont {E.~W.~N.}\ \bibnamefont
  {Glover}}, \bibinfo {author} {\bibfnamefont {T.}~\bibnamefont {Huber}},
  \bibinfo {author} {\bibfnamefont {N.}~\bibnamefont {Ikizlerli}}, \ and\
  \bibinfo {author} {\bibfnamefont {C.}~\bibnamefont {Studerus}},\ }\href
  {\doibase 10.1007/JHEP06(2010)094} {\bibfield  {journal} {\bibinfo  {journal}
  {JHEP}\ }\textbf {\bibinfo {volume} {06}},\ \bibinfo {pages} {094} (\bibinfo
  {year} {2010})},\ \Eprint {http://arxiv.org/abs/1004.3653} {arXiv:1004.3653
  [hep-ph]} \BibitemShut {NoStop}%
\bibitem [{\citenamefont {Gehrmann-De~Ridder}\ \emph
  {et~al.}(2018)\citenamefont {Gehrmann-De~Ridder}, \citenamefont {Gehrmann},
  \citenamefont {Glover}, \citenamefont {Huss},\ and\ \citenamefont
  {Walker}}]{Gehrmann-DeRidder:2017mvr}%
  \BibitemOpen
  \bibfield  {author} {\bibinfo {author} {\bibfnamefont {A.}~\bibnamefont
  {Gehrmann-De~Ridder}}, \bibinfo {author} {\bibfnamefont {T.}~\bibnamefont
  {Gehrmann}}, \bibinfo {author} {\bibfnamefont {E.~W.~N.}\ \bibnamefont
  {Glover}}, \bibinfo {author} {\bibfnamefont {A.}~\bibnamefont {Huss}}, \ and\
  \bibinfo {author} {\bibfnamefont {D.~M.}\ \bibnamefont {Walker}},\ }\href
  {\doibase 10.1103/PhysRevLett.120.122001} {\bibfield  {journal} {\bibinfo
  {journal} {Phys. Rev. Lett.}\ }\textbf {\bibinfo {volume} {120}},\ \bibinfo
  {pages} {122001} (\bibinfo {year} {2018})},\ \Eprint
  {http://arxiv.org/abs/1712.07543} {arXiv:1712.07543 [hep-ph]} \BibitemShut
  {NoStop}%
\bibitem [{\citenamefont {Gehrmann-De~Ridder}\ \emph
  {et~al.}(2019)\citenamefont {Gehrmann-De~Ridder}, \citenamefont {Gehrmann},
  \citenamefont {Glover}, \citenamefont {Huss},\ and\ \citenamefont
  {Walker}}]{Gehrmann-DeRidder:2019avi}%
  \BibitemOpen
  \bibfield  {author} {\bibinfo {author} {\bibfnamefont {A.}~\bibnamefont
  {Gehrmann-De~Ridder}}, \bibinfo {author} {\bibfnamefont {T.}~\bibnamefont
  {Gehrmann}}, \bibinfo {author} {\bibfnamefont {E.~W.~N.}\ \bibnamefont
  {Glover}}, \bibinfo {author} {\bibfnamefont {A.}~\bibnamefont {Huss}}, \ and\
  \bibinfo {author} {\bibfnamefont {D.~M.}\ \bibnamefont {Walker}},\ }\href
  {\doibase 10.1140/epjc/s10052-019-7010-2} {\bibfield  {journal} {\bibinfo
  {journal} {Eur. Phys. J. C}\ }\textbf {\bibinfo {volume} {79}},\ \bibinfo
  {pages} {526} (\bibinfo {year} {2019})},\ \Eprint
  {http://arxiv.org/abs/1901.11041} {arXiv:1901.11041 [hep-ph]} \BibitemShut
  {NoStop}%
\bibitem [{\citenamefont {Gehrmann-De~Ridder}\ \emph
  {et~al.}(2005)\citenamefont {Gehrmann-De~Ridder}, \citenamefont {Gehrmann},\
  and\ \citenamefont {Glover}}]{hep-ph/0505111}%
  \BibitemOpen
  \bibfield  {author} {\bibinfo {author} {\bibfnamefont {A.}~\bibnamefont
  {Gehrmann-De~Ridder}}, \bibinfo {author} {\bibfnamefont {T.}~\bibnamefont
  {Gehrmann}}, \ and\ \bibinfo {author} {\bibfnamefont {E.~W.~N.}\ \bibnamefont
  {Glover}},\ }\href {\doibase 10.1088/1126-6708/2005/09/056} {\bibfield
  {journal} {\bibinfo  {journal} {JHEP}\ }\textbf {\bibinfo {volume} {09}},\
  \bibinfo {pages} {056} (\bibinfo {year} {2005})},\ \Eprint
  {http://arxiv.org/abs/hep-ph/0505111} {arXiv:hep-ph/0505111} \BibitemShut
  {NoStop}%
\bibitem [{\citenamefont {Daleo}\ \emph {et~al.}(2007)\citenamefont {Daleo},
  \citenamefont {Gehrmann},\ and\ \citenamefont {Maitre}}]{hep-ph/0612257}%
  \BibitemOpen
  \bibfield  {author} {\bibinfo {author} {\bibfnamefont {A.}~\bibnamefont
  {Daleo}}, \bibinfo {author} {\bibfnamefont {T.}~\bibnamefont {Gehrmann}}, \
  and\ \bibinfo {author} {\bibfnamefont {D.}~\bibnamefont {Maitre}},\ }\href
  {\doibase 10.1088/1126-6708/2007/04/016} {\bibfield  {journal} {\bibinfo
  {journal} {JHEP}\ }\textbf {\bibinfo {volume} {04}},\ \bibinfo {pages} {016}
  (\bibinfo {year} {2007})},\ \Eprint {http://arxiv.org/abs/hep-ph/0612257}
  {arXiv:hep-ph/0612257} \BibitemShut {NoStop}%
\bibitem [{\citenamefont {Currie}\ \emph {et~al.}(2013)\citenamefont {Currie},
  \citenamefont {Glover},\ and\ \citenamefont {Wells}}]{1301.4693}%
  \BibitemOpen
  \bibfield  {author} {\bibinfo {author} {\bibfnamefont {J.}~\bibnamefont
  {Currie}}, \bibinfo {author} {\bibfnamefont {E.~W.~N.}\ \bibnamefont
  {Glover}}, \ and\ \bibinfo {author} {\bibfnamefont {S.}~\bibnamefont
  {Wells}},\ }\href {\doibase 10.1007/JHEP04(2013)066} {\bibfield  {journal}
  {\bibinfo  {journal} {JHEP}\ }\textbf {\bibinfo {volume} {04}},\ \bibinfo
  {pages} {066} (\bibinfo {year} {2013})},\ \Eprint
  {http://arxiv.org/abs/1301.4693} {arXiv:1301.4693 [hep-ph]} \BibitemShut
  {NoStop}%
\bibitem [{\citenamefont {Zyla}\ \emph {et~al.}(2020)\citenamefont {Zyla} \emph
  {et~al.}}]{ParticleDataGroup:2020ssz}%
  \BibitemOpen
  \bibfield  {author} {\bibinfo {author} {\bibfnamefont {P.~A.}\ \bibnamefont
  {Zyla}} \emph {et~al.} (\bibinfo {collaboration} {Particle Data Group}),\
  }\href {\doibase 10.1093/ptep/ptaa104} {\bibfield  {journal} {\bibinfo
  {journal} {PTEP}\ }\textbf {\bibinfo {volume} {2020}},\ \bibinfo {pages}
  {083C01} (\bibinfo {year} {2020})}\BibitemShut {NoStop}%
\bibitem [{\citenamefont {Ball}\ \emph {et~al.}(2017)\citenamefont {Ball} \emph
  {et~al.}}]{NNPDF:2017mvq}%
  \BibitemOpen
  \bibfield  {author} {\bibinfo {author} {\bibfnamefont {R.~D.}\ \bibnamefont
  {Ball}} \emph {et~al.} (\bibinfo {collaboration} {NNPDF}),\ }\href {\doibase
  10.1140/epjc/s10052-017-5199-5} {\bibfield  {journal} {\bibinfo  {journal}
  {Eur. Phys. J. C}\ }\textbf {\bibinfo {volume} {77}},\ \bibinfo {pages} {663}
  (\bibinfo {year} {2017})},\ \Eprint {http://arxiv.org/abs/1706.00428}
  {arXiv:1706.00428 [hep-ph]} \BibitemShut {NoStop}%
\bibitem [{\citenamefont {Ball}\ \emph {et~al.}(2022)\citenamefont {Ball} \emph
  {et~al.}}]{NNPDF:2021njg}%
  \BibitemOpen
  \bibfield  {author} {\bibinfo {author} {\bibfnamefont {R.~D.}\ \bibnamefont
  {Ball}} \emph {et~al.} (\bibinfo {collaboration} {NNPDF}),\ }\href {\doibase
  10.1140/epjc/s10052-022-10328-7} {\bibfield  {journal} {\bibinfo  {journal}
  {Eur. Phys. J. C}\ }\textbf {\bibinfo {volume} {82}},\ \bibinfo {pages} {428}
  (\bibinfo {year} {2022})},\ \Eprint {http://arxiv.org/abs/2109.02653}
  {arXiv:2109.02653 [hep-ph]} \BibitemShut {NoStop}%
\bibitem [{\citenamefont {Buckley}\ \emph {et~al.}(2015)\citenamefont
  {Buckley}, \citenamefont {Ferrando}, \citenamefont {Lloyd}, \citenamefont
  {Nordstr\"om}, \citenamefont {Page}, \citenamefont {R\"ufenacht},
  \citenamefont {Sch\"onherr},\ and\ \citenamefont {Watt}}]{Buckley:2014ana}%
  \BibitemOpen
  \bibfield  {author} {\bibinfo {author} {\bibfnamefont {A.}~\bibnamefont
  {Buckley}}, \bibinfo {author} {\bibfnamefont {J.}~\bibnamefont {Ferrando}},
  \bibinfo {author} {\bibfnamefont {S.}~\bibnamefont {Lloyd}}, \bibinfo
  {author} {\bibfnamefont {K.}~\bibnamefont {Nordstr\"om}}, \bibinfo {author}
  {\bibfnamefont {B.}~\bibnamefont {Page}}, \bibinfo {author} {\bibfnamefont
  {M.}~\bibnamefont {R\"ufenacht}}, \bibinfo {author} {\bibfnamefont
  {M.}~\bibnamefont {Sch\"onherr}}, \ and\ \bibinfo {author} {\bibfnamefont
  {G.}~\bibnamefont {Watt}},\ }\href {\doibase 10.1140/epjc/s10052-015-3318-8}
  {\bibfield  {journal} {\bibinfo  {journal} {Eur. Phys. J. C}\ }\textbf
  {\bibinfo {volume} {75}},\ \bibinfo {pages} {132} (\bibinfo {year} {2015})},\
  \Eprint {http://arxiv.org/abs/1412.7420} {arXiv:1412.7420 [hep-ph]}
  \BibitemShut {NoStop}%
\bibitem [{\citenamefont {Duhr}\ \emph
  {et~al.}(2020{\natexlab{b}})\citenamefont {Duhr}, \citenamefont {Dulat},\
  and\ \citenamefont {Mistlberger}}]{Duhr:2020seh}%
  \BibitemOpen
  \bibfield  {author} {\bibinfo {author} {\bibfnamefont {C.}~\bibnamefont
  {Duhr}}, \bibinfo {author} {\bibfnamefont {F.}~\bibnamefont {Dulat}}, \ and\
  \bibinfo {author} {\bibfnamefont {B.}~\bibnamefont {Mistlberger}},\ }\href
  {\doibase 10.1103/PhysRevLett.125.172001} {\bibfield  {journal} {\bibinfo
  {journal} {Phys. Rev. Lett.}\ }\textbf {\bibinfo {volume} {125}},\ \bibinfo
  {pages} {172001} (\bibinfo {year} {2020}{\natexlab{b}})},\ \Eprint
  {http://arxiv.org/abs/2001.07717} {arXiv:2001.07717 [hep-ph]} \BibitemShut
  {NoStop}%
\bibitem [{\citenamefont {Duhr}\ and\ \citenamefont
  {Mistlberger}(2022)}]{Duhr:2021vwj}%
  \BibitemOpen
  \bibfield  {author} {\bibinfo {author} {\bibfnamefont {C.}~\bibnamefont
  {Duhr}}\ and\ \bibinfo {author} {\bibfnamefont {B.}~\bibnamefont
  {Mistlberger}},\ }\href {\doibase 10.1007/JHEP03(2022)116} {\bibfield
  {journal} {\bibinfo  {journal} {JHEP}\ }\textbf {\bibinfo {volume} {03}},\
  \bibinfo {pages} {116} (\bibinfo {year} {2022})},\ \Eprint
  {http://arxiv.org/abs/2111.10379} {arXiv:2111.10379 [hep-ph]} \BibitemShut
  {NoStop}%
\bibitem [{\citenamefont {Chen}\ \emph
  {et~al.}(2022{\natexlab{b}})\citenamefont {Chen}, \citenamefont {Gehrmann},
  \citenamefont {Glover}, \citenamefont {Huss}, \citenamefont {Monni},
  \citenamefont {Re}, \citenamefont {Rottoli},\ and\ \citenamefont
  {Torrielli}}]{Chen:2022cgv}%
  \BibitemOpen
  \bibfield  {author} {\bibinfo {author} {\bibfnamefont {X.}~\bibnamefont
  {Chen}}, \bibinfo {author} {\bibfnamefont {T.}~\bibnamefont {Gehrmann}},
  \bibinfo {author} {\bibfnamefont {E.~W.~N.}\ \bibnamefont {Glover}}, \bibinfo
  {author} {\bibfnamefont {A.}~\bibnamefont {Huss}}, \bibinfo {author}
  {\bibfnamefont {P.~F.}\ \bibnamefont {Monni}}, \bibinfo {author}
  {\bibfnamefont {E.}~\bibnamefont {Re}}, \bibinfo {author} {\bibfnamefont
  {L.}~\bibnamefont {Rottoli}}, \ and\ \bibinfo {author} {\bibfnamefont
  {P.}~\bibnamefont {Torrielli}},\ }\href {\doibase
  10.1103/PhysRevLett.128.252001} {\bibfield  {journal} {\bibinfo  {journal}
  {Phys. Rev. Lett.}\ }\textbf {\bibinfo {volume} {128}},\ \bibinfo {pages}
  {252001} (\bibinfo {year} {2022}{\natexlab{b}})},\ \Eprint
  {http://arxiv.org/abs/2203.01565} {arXiv:2203.01565 [hep-ph]} \BibitemShut
  {NoStop}%
\bibitem [{\citenamefont {Aaltonen}\ \emph {et~al.}(2009)\citenamefont
  {Aaltonen} \emph {et~al.}}]{CDF:2009cjw}%
  \BibitemOpen
  \bibfield  {author} {\bibinfo {author} {\bibfnamefont {T.}~\bibnamefont
  {Aaltonen}} \emph {et~al.} (\bibinfo {collaboration} {CDF}),\ }\href
  {\doibase 10.1103/PhysRevLett.102.181801} {\bibfield  {journal} {\bibinfo
  {journal} {Phys. Rev. Lett.}\ }\textbf {\bibinfo {volume} {102}},\ \bibinfo
  {pages} {181801} (\bibinfo {year} {2009})},\ \Eprint
  {http://arxiv.org/abs/0901.2169} {arXiv:0901.2169 [hep-ex]} \BibitemShut
  {NoStop}%
\bibitem [{\citenamefont {Abazov}\ \emph {et~al.}(2014)\citenamefont {Abazov}
  \emph {et~al.}}]{D0:2013lql}%
  \BibitemOpen
  \bibfield  {author} {\bibinfo {author} {\bibfnamefont {V.~M.}\ \bibnamefont
  {Abazov}} \emph {et~al.} (\bibinfo {collaboration} {D0}),\ }\href {\doibase
  10.1103/PhysRevLett.112.151803} {\bibfield  {journal} {\bibinfo  {journal}
  {Phys. Rev. Lett.}\ }\textbf {\bibinfo {volume} {112}},\ \bibinfo {pages}
  {151803} (\bibinfo {year} {2014})},\ \bibinfo {note} {[Erratum:
  Phys.Rev.Lett. 114, 049901 (2015)]},\ \Eprint
  {http://arxiv.org/abs/1312.2895} {arXiv:1312.2895 [hep-ex]} \BibitemShut
  {NoStop}%
\bibitem [{\citenamefont {Aad}\ \emph {et~al.}(2011)\citenamefont {Aad} \emph
  {et~al.}}]{ATLAS:2011pph}%
  \BibitemOpen
  \bibfield  {author} {\bibinfo {author} {\bibfnamefont {G.}~\bibnamefont
  {Aad}} \emph {et~al.} (\bibinfo {collaboration} {ATLAS}),\ }\href {\doibase
  10.1016/j.physletb.2011.05.024} {\bibfield  {journal} {\bibinfo  {journal}
  {Phys. Lett. B}\ }\textbf {\bibinfo {volume} {701}},\ \bibinfo {pages} {31}
  (\bibinfo {year} {2011})},\ \Eprint {http://arxiv.org/abs/1103.2929}
  {arXiv:1103.2929 [hep-ex]} \BibitemShut {NoStop}%
\bibitem [{\citenamefont {Aaltonen}\ \emph {et~al.}(2008)\citenamefont
  {Aaltonen} \emph {et~al.}}]{CDF:2007tdb}%
  \BibitemOpen
  \bibfield  {author} {\bibinfo {author} {\bibfnamefont {T.}~\bibnamefont
  {Aaltonen}} \emph {et~al.} (\bibinfo {collaboration} {CDF}),\ }\href
  {\doibase 10.1103/PhysRevLett.100.071801} {\bibfield  {journal} {\bibinfo
  {journal} {Phys. Rev. Lett.}\ }\textbf {\bibinfo {volume} {100}},\ \bibinfo
  {pages} {071801} (\bibinfo {year} {2008})},\ \Eprint
  {http://arxiv.org/abs/0710.4112} {arXiv:0710.4112 [hep-ex]} \BibitemShut
  {NoStop}%
\bibitem [{\citenamefont {Abazov}\ \emph {et~al.}(2009)\citenamefont {Abazov}
  \emph {et~al.}}]{D0:2009oet}%
  \BibitemOpen
  \bibfield  {author} {\bibinfo {author} {\bibfnamefont {V.~M.}\ \bibnamefont
  {Abazov}} \emph {et~al.} (\bibinfo {collaboration} {D0}),\ }\href {\doibase
  10.1103/PhysRevLett.103.231802} {\bibfield  {journal} {\bibinfo  {journal}
  {Phys. Rev. Lett.}\ }\textbf {\bibinfo {volume} {103}},\ \bibinfo {pages}
  {231802} (\bibinfo {year} {2009})},\ \Eprint {http://arxiv.org/abs/0909.4814}
  {arXiv:0909.4814 [hep-ex]} \BibitemShut {NoStop}%
\bibitem [{\citenamefont {Isaacson}\ \emph {et~al.}(2022)\citenamefont
  {Isaacson}, \citenamefont {Fu},\ and\ \citenamefont
  {Yuan}}]{Isaacson:2022rts}%
  \BibitemOpen
  \bibfield  {author} {\bibinfo {author} {\bibfnamefont {J.}~\bibnamefont
  {Isaacson}}, \bibinfo {author} {\bibfnamefont {Y.}~\bibnamefont {Fu}}, \ and\
  \bibinfo {author} {\bibfnamefont {C.~P.}\ \bibnamefont {Yuan}},\ }\href@noop
  {} {\  (\bibinfo {year} {2022})},\ \Eprint {http://arxiv.org/abs/2205.02788}
  {arXiv:2205.02788 [hep-ph]} \BibitemShut {NoStop}%
\bibitem [{\citenamefont {Gao}\ \emph {et~al.}(2022)\citenamefont {Gao},
  \citenamefont {Liu},\ and\ \citenamefont {Xie}}]{Gao:2022wxk}%
  \BibitemOpen
  \bibfield  {author} {\bibinfo {author} {\bibfnamefont {J.}~\bibnamefont
  {Gao}}, \bibinfo {author} {\bibfnamefont {D.}~\bibnamefont {Liu}}, \ and\
  \bibinfo {author} {\bibfnamefont {K.}~\bibnamefont {Xie}},\ }\href {\doibase
  10.1088/1674-1137/ac930b} {\bibfield  {journal} {\bibinfo  {journal} {Chin.
  Phys. C}\ }\textbf {\bibinfo {volume} {46}},\ \bibinfo {pages} {123110}
  (\bibinfo {year} {2022})},\ \Eprint {http://arxiv.org/abs/2205.03942}
  {arXiv:2205.03942 [hep-ph]} \BibitemShut {NoStop}%
\bibitem [{\citenamefont {Camarda}\ \emph {et~al.}(2022)\citenamefont
  {Camarda}, \citenamefont {Cieri},\ and\ \citenamefont
  {Ferrera}}]{Camarda:2021jsw}%
  \BibitemOpen
  \bibfield  {author} {\bibinfo {author} {\bibfnamefont {S.}~\bibnamefont
  {Camarda}}, \bibinfo {author} {\bibfnamefont {L.}~\bibnamefont {Cieri}}, \
  and\ \bibinfo {author} {\bibfnamefont {G.}~\bibnamefont {Ferrera}},\ }\href
  {\doibase 10.1140/epjc/s10052-022-10510-x} {\bibfield  {journal} {\bibinfo
  {journal} {Eur. Phys. J. C}\ }\textbf {\bibinfo {volume} {82}},\ \bibinfo
  {pages} {575} (\bibinfo {year} {2022})},\ \Eprint
  {http://arxiv.org/abs/2111.14509} {arXiv:2111.14509 [hep-ph]} \BibitemShut
  {NoStop}%
\bibitem [{\citenamefont {Buonocore}\ \emph {et~al.}(2022)\citenamefont
  {Buonocore}, \citenamefont {Kallweit}, \citenamefont {Rottoli},\ and\
  \citenamefont {Wiesemann}}]{Buonocore:2021tke}%
  \BibitemOpen
  \bibfield  {author} {\bibinfo {author} {\bibfnamefont {L.}~\bibnamefont
  {Buonocore}}, \bibinfo {author} {\bibfnamefont {S.}~\bibnamefont {Kallweit}},
  \bibinfo {author} {\bibfnamefont {L.}~\bibnamefont {Rottoli}}, \ and\
  \bibinfo {author} {\bibfnamefont {M.}~\bibnamefont {Wiesemann}},\ }\href
  {\doibase 10.1016/j.physletb.2022.137118} {\bibfield  {journal} {\bibinfo
  {journal} {Phys. Lett. B}\ }\textbf {\bibinfo {volume} {829}},\ \bibinfo
  {pages} {137118} (\bibinfo {year} {2022})},\ \Eprint
  {http://arxiv.org/abs/2111.13661} {arXiv:2111.13661 [hep-ph]} \BibitemShut
  {NoStop}%
\bibitem [{\citenamefont {Achard}\ \emph {et~al.}(2006)\citenamefont {Achard}
  \emph {et~al.}}]{L3:2005fft}%
  \BibitemOpen
  \bibfield  {author} {\bibinfo {author} {\bibfnamefont {P.}~\bibnamefont
  {Achard}} \emph {et~al.} (\bibinfo {collaboration} {L3}),\ }\href {\doibase
  10.1140/epjc/s2005-02459-6} {\bibfield  {journal} {\bibinfo  {journal} {Eur.
  Phys. J. C}\ }\textbf {\bibinfo {volume} {45}},\ \bibinfo {pages} {569}
  (\bibinfo {year} {2006})},\ \Eprint {http://arxiv.org/abs/hep-ex/0511049}
  {arXiv:hep-ex/0511049} \BibitemShut {NoStop}%
\end{thebibliography}%

\end{document}